\pdfoutput=1

\documentclass[twocolumn]{aastex62}
\usepackage{multirow}

\shorttitle{Physical properties of HSC--FIRST radio galaxies}
\shortauthors{Toba et al.}

\begin{document}

\title{A Wide and Deep Exploration of Radio Galaxies with Subaru HSC (WERGS). \\
II. Physical Properties derived from the SED Fitting with Optical, Infrared, and Radio Data}

\correspondingauthor{Yoshiki Toba}
\email{toba@kusastro.kyoto-u.ac.jp}

\author[0000-0002-3531-7863]{Yoshiki Toba}
\affil{Department of Astronomy, Kyoto University, Kitashirakawa-Oiwake-cho, Sakyo-ku, Kyoto 606-8502, Japan}
\affil{Academia Sinica Institute of Astronomy and Astrophysics, 11F of Astronomy-Mathematics Building, AS/NTU, No.1, Section 4, Roosevelt Road, Taipei 10617, Taiwan}
\affil{Research Center for Space and Cosmic Evolution, Ehime University, 2-5 Bunkyo-cho, Matsuyama, Ehime 790-8577, Japan}

\author{Takuji Yamashita}
\affil{Research Center for Space and Cosmic Evolution, Ehime University, 2-5 Bunkyo-cho, Matsuyama, Ehime 790-8577, Japan}
\affil{National Astronomical Observatory of Japan, 2-21-1 Osawa, Mitaka, Tokyo 181-8588, Japan}

\author{Tohru Nagao}
\affil{Research Center for Space and Cosmic Evolution, Ehime University, 2-5 Bunkyo-cho, Matsuyama, Ehime 790-8577, Japan}

\author{Wei-Hao Wang}
\affil{Academia Sinica Institute of Astronomy and Astrophysics, 11F of Astronomy-Mathematics Building, AS/NTU, No.1, Section 4, Roosevelt Road, Taipei 10617, Taiwan}

\author{Yoshihiro Ueda}
\affil{Department of Astronomy, Kyoto University, Kitashirakawa-Oiwake-cho, Sakyo-ku, Kyoto 606-8502, Japan}

\author{Kohei Ichikawa}
\affil{Astronomical Institute, Tohoku University, 6-3 Aramaki, Aoba-ku, Sendai, Miyagi 980-8578, Japan}
\affil{Frontier Research Institute for Interdisciplinary Sciences, Tohoku University, 6-3 Aramaki, Aoba-ku, Sendai, Miyagi 980-8578, Japan}

\author{Toshihiro Kawaguchi}
\affil{Department of Economics, Management and Information Science, Onomichi City University, Hisayamada 1600-2, Onomichi, Hiroshima 722-8506, Japan}

\author{Masayuki Akiyama}
\affil{Astronomical Institute, Tohoku University, 6-3 Aramaki, Aoba-ku, Sendai, Miyagi 980-8578, Japan}

\author{Bau-Ching Hsieh}
\affil{Academia Sinica Institute of Astronomy and Astrophysics, 11F of Astronomy-Mathematics Building, AS/NTU, No.1, Section 4, Roosevelt Road, Taipei 10617, Taiwan}

\author{Masaru Kajisawa}
\affil{Graduate School of Science and Engineering, Ehime University, Bunkyo-cho, Matsuyama, Ehime 790-8577, Japan}
\affil{Research Center for Space and Cosmic Evolution, Ehime University, 2-5 Bunkyo-cho, Matsuyama, Ehime 790-8577, Japan}

\author{Chien-Hsiu Lee}
\affil{National Optical Astronomy Observatory, 950 N Cherry Ave., Tucson, AZ 85719, USA}

\author{Yoshiki Matsuoka}
\affil{Research Center for Space and Cosmic Evolution, Ehime University, 2-5 Bunkyo-cho, Matsuyama, Ehime 790-8577, Japan}

\author{Akatoki Noboriguchi}
\affil{Graduate School of Science and Engineering, Ehime University, Bunkyo-cho, Matsuyama, Ehime 790-8577, Japan}

\author{Masafusa Onoue}
\affiliation{Max-Planck-Institut f\"ur Astronomie, K\"onigstuhl 17, D-69117 Heidelberg, Germany}

\author{Malte Schramm}
\affil{National Astronomical Observatory of Japan, 2-21-1 Osawa, Mitaka, Tokyo 181-8588, Japan}

\author{Masayuki Tanaka}
\affil{National Astronomical Observatory of Japan, 2-21-1 Osawa, Mitaka, Tokyo 181-8588, Japan}
\affil{Department of Astronomy, School of Science, Graduate University for Advanced Studies (SOKENDAI), 2-21-1 Osawa, Mitaka, Tokyo 181-8588, Japan}

\author{Yutaka Komiyama}
\affil{National Astronomical Observatory of Japan, 2-21-1 Osawa, Mitaka, Tokyo 181-8588, Japan}
\affil{Department of Astronomy, School of Science, Graduate University for Advanced Studies (SOKENDAI), 2-21-1 Osawa, Mitaka, Tokyo 181-8588, Japan}




\begin{abstract}
We present physical properties of radio galaxies (RGs) with $f_{\rm 1.4 GHz} >$ 1 mJy discovered by Subaru Hyper Supreme-Cam (HSC) and VLA Faint Images of the Radio Sky at Twenty-Centimeters (FIRST) survey. 
For 1056 FIRST RGs at $0 < z \leq 1.7$ with HSC counterparts in about 100 deg$^2$, we compiled multi-wavelength data of optical, near-infrared (IR), mid-IR, far-IR, and radio (150 MHz).
We derived their color excess ($E (B-V)_{*}$), stellar mass, star formation rate (SFR), IR luminosity, the ratio of IR and radio luminosity ($q_{\rm IR}$), and radio spectral index ($\alpha_{\rm radio}$) that are derived from the SED fitting with {\tt CIGALE}.
We also estimated Eddington ratio based on stellar mass and integration of the best-fit SEDs of AGN component.
We found that $E (B-V)_{*}$, SFR, and IR luminosity clearly depend on redshift while stellar mass, $q_{\rm IR}$, and $\alpha_{\rm radio}$ do not significantly depend on redshift.
Since optically-faint ($i_{\rm AB} \geq 21.3$) RGs that are newly discovered by our RG survey tend to be high redshift, they tend to not only have a large dust extinction and low stellar mass but also have high SFR and AGN luminosity, high IR luminosity, and high Eddington ratio compared to optically-bright ones.
The physical properties of a fraction of RGs in our sample seem to differ from a classical view of RGs with massive stellar mass, low SFR, and low Eddington ratio, demonstrating that our RG survey with HSC and FIRST provides us curious RGs among entire RG population.
\end{abstract}



\section{Introduction} 
\label{intro}

In the last decade, observational and theoretical works have reported that feedback from radio active galactic nuclei (AGNs) harbored in radio galaxies (RGs) and radio-loud quasars can play an important role in the formation and evolution of galaxies \citep[e.g.,][]{Croton_06,Fabian_12}.  
Mechanical injection of energy from RGs provides an impact on the gas reservoirs in galaxies and galaxy clusters \citep{Morganti_13}. 
Such AGN feedback could regulate star formation (SF) and even the growth of supermassive black holes (SMBHs) in galaxies. 
Therefore, it is important to investigate the physical properties related to SF and AGN activity for RGs as a function of redshift in order to understand a full picture of the formation and evolution of galaxies.

Multi-wavelength dataset of optical and infrared (IR) for RGs is crucial for studying their physical properties such as stellar mass, AGN/SF activity, and star formation rate (SFR). 
For example, a combination of  National Radio Astronomy Observatory (NRAO) Very Large Array (VLA) Sky Survey \citep[NVSS; ][]{Condon} or the VLA Faint Images of the Radio Sky at Twenty-Centimeters survey \citep[FIRST; ][]{Becker}, and the Sloan Digital Sky Survey \citep[SDSS; ][]{York} provided a lager number of RGs with optical counterparts in the local Universe \citep{Ivezic,Best_05,Helfand}, allowing us a statical investigation of those ``optically bright'' RGs with $r < 22.2$ mag at redshift $z < 0.5$.
These objects have been well studied in terms of UV/optical properties \citep[e.g.,][]{de_Ruiter}, morphologies \citep[e.g.,][]{Liske,Aniyan,Lukic}, mid-IR (MIR) properties \citep[e.g.,][]{Gurkan_14}, and far-IR (FIR) properties \citep[e.g.,][]{Gurkan_15,Gurkan_18} as well as black hole (BH) mass and its accretion rate \citep[e.g.,][]{Best_12}. 
Almost all of the optically bright local RGs have elliptical hosts with stellar mass of $> 10^{10.5}$ M$_{\sun}$ and SFR of $<$ 10 M$_{\sun}$ yr$^{-1}$ \citep{Best_12}.
Only a small fraction of the local RGs has relatively small stellar mass with moderate star-forming activities \citep{Smolcic_09,Best_12}.

At the high-$z$ Universe ($z>1$), known RGs are powerful or radio-luminous ($L_{\rm radio} \gtrsim 10^{26}$~W~Hz$^{-1}$, corresponding to $>0.1$~mJy).
The powerful high-$z$ RGs are dominated by the evolved stellar populations with a stellar mass of $10^{11-12}~M_{\sun}$ \citep[e.g., ][]{Rocca-Volmerange_04,Seymour_07,Casey}.
The IR luminosity ($L_{\rm IR}$) of those powerful high-$z$ RGs often exceed $10^{12}\, L_{\sun}$ that is classified as ultraluminous IR galaxy \citep[ULIRG; ][]{Sanders}.
They also show the evidence of high SFR and high BH accretion rate through IR and sub-millimeter observations \citep[e.g.,][]{Chapman,Magnelli,Seymour_12,Drouart_14,Bonzini_15}.
On the other hand, \cite{Falkendal} investigated the SFR of those powerful high-$z$ RGs based on multi-wavelength SEDs with taking into account their synchrotron emission.
They reported that their SFRs are indeed lower than those of a main sequence of galaxies, suggesting an importance of multi-wavelength analysis for RGs.

Deep radio and optical observations enable us to find much more fainter RGs \citep[see ][and references therein]{Padovani} and to provide a comprehensive understanding by connecting RGs between local and high-$z$ Universe.
\citet{Delvecchio_18} investigated RGs in the VLA-COSMOS field \citep{Smolcic_17b} based on a multi-wavelength dataset \citep{Smolcic_17a,Laigle_16}.
They found that an average BH mass accretion rate, represented by a ratio of bolometric luminosity to stellar mass, increases with increasing redshift up to $z\sim 4$. 
They also reported that this trend is similar to a fact that fraction of star-forming host galaxies also increases with increasing redshift.
Although their statistical experiment was performed with a relatively small area ($\sim 2$~deg$^2$), a wide-field survey with deep radio and optical facilities enables to find a large number of ``optically faint'' RGs, providing us a laboratory to investigate their evolution in more high resolutions in redshifts and luminosities.

Recently, \citet[][Paper I]{Yamashita} performed a systematic search for RGs and quasars as a project, so-called ``the Wide and Deep Exploration of Radio Galaxies with Subaru HSC (WERGS).''
They reported the result of optical identifications of radio sources detected by FIRST with the Hyper Suprime-Cam \citep[HSC; ][]{Miyazaki_12,Miyazaki} \citep
[see also][]{Furusawa,Kawanomoto,Komiyama} Subaru Strategic Program survey \citep[HSC-SSP; ][]{Aihara_18a}.
By cross-matching the final data release of the FIRST survey \citep{Helfand} with HSC S16A data \citep{Aihara_18b}, they found 3579 optical counterparts of FIRST sources in a 154 deg$^2$ of a HSC-SSP Wide field (see Section \ref{SSS}).
Their radio flux densities at 1.4 GHz (20 cm) are above 1 mJy while about 60\% of them are optically-faint ones with $i \geq$ 21.3 mag that are undetected by the SDSS, allowing us to explore a new parameter space, i.e., optically-faint bright radio sources.
Plenty of RG and quasar sample also gives an opportunity to discover a specially rare population, for example, a RG at high redshift (Yamashita et al. in preparation) and extremely radio-loud quasars (Ichikawa et al. in preparation).

This is the second in a series of papers from the WERGS project, in which we report the physical properties of radio-loud galaxies at $0 < z \leq 1.7$ with $i$-band magnitude between 18 and 26, that are derived from the Spectral Energy Distribution (SED) fitting of multi-wavelength data.
In this paper, we follow the same definition of RGs and quasars as adopted in \cite{Yamashita}.
But we removed stellar objects, i.e., radio-loud quasars that are optically unresolved objects based on optical morphology \citep[see][]{Yamashita}, and focus only on RGs that have optically resolved morphologies.

The structure of this paper is as follows. 
Section \ref{DA} describes the sample selection of RGs, the multi-wavelength dataset, and our SED modeling. 
In Section \ref{result}, we report the result of SED fitting and the derived physical quantities of RGs detected by the HSC and FIRST.
In Section \ref{discussion}, we discuss a possible selection bias, an uncertainty of our SED fitting, and BH mass accretion rate for our sample.
We summarize this work in Section \ref{summary}. 
All information about our RG sample such as coordinates, multi-band photometry, derived physical quantities are available as a catalog (see Appendix \ref{app1}). 
We also provide best-fit SED templates of those RGs (see Appendix \ref{app2}).  
Throughout this paper, the adopted cosmology is a flat universe with $H_0$ = 70 km s$^{-1}$ Mpc$^{-1}$, $\Omega_{\rm M}$ = 0.27, and $\Omega_{\Lambda}$ = 0.73, that are same as those adopted in\cite{Yamashita}.
Unless otherwise noted, all magnitudes refer to the AB system.

\section{Data and analysis} 
\label{DA}
\begin{figure}
\plotone{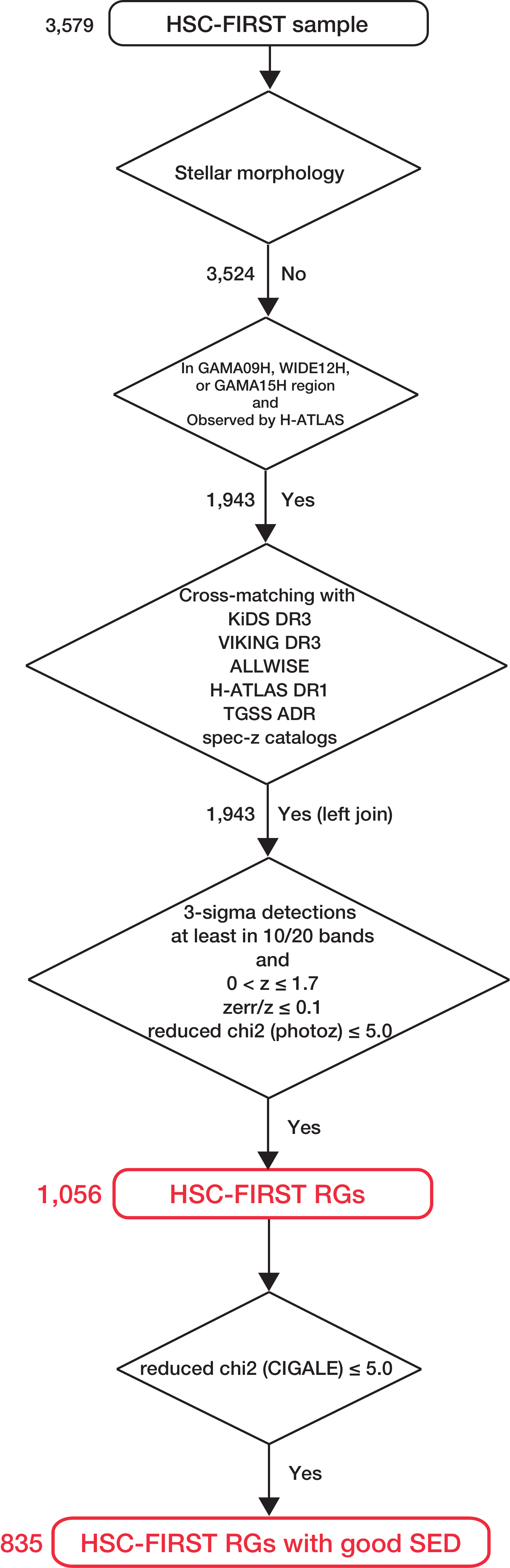}
\caption{Flow chart of the sample selection process.}
\label{SS}
\end{figure}
\begin{figure*}
\plotone{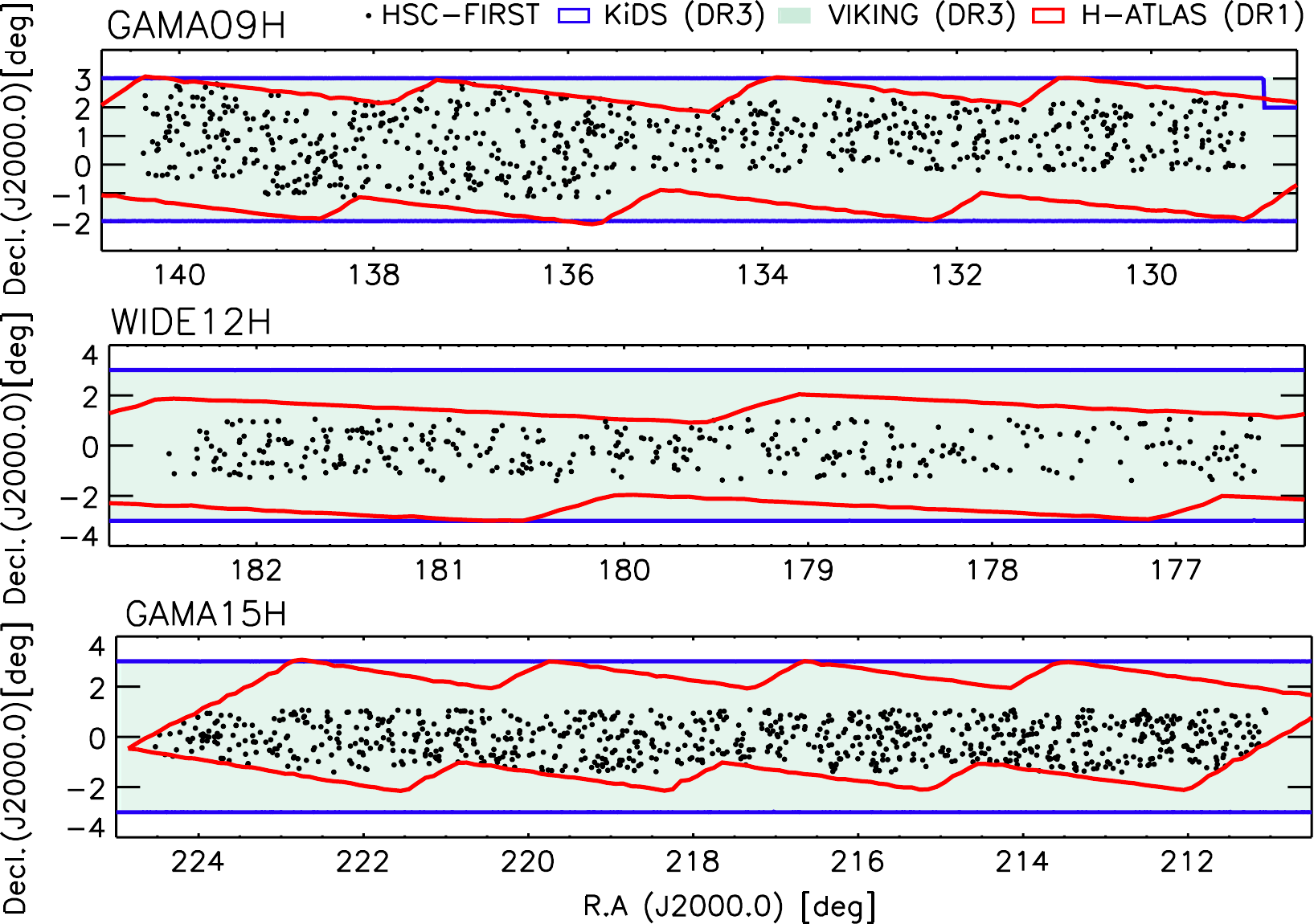}
\caption{Spatial distribution (J2000.0) of 1943 HSC--FIRST radio galaxies (black pointes) in GAMA09H (top), WIDE12 (middle). and GAMA15H (bottom) field. Blue, green, and red squares represent survey footprint of KiDS, VIKING, and H-ATLAS, respectively. Those regions are completely covered by ALLWISE and TGSS. There are 754, 344, and 845 HSC--FIRST objects in the GAMA09H, WIDE12, and GAMA15H, respectively.}
\label{region}
\end{figure*}

\subsection{Sample selection}
\label{SSS}

Figure \ref{SS} shows a flow chart of our sample selection process.
The original sample was drawn from 3579 RGs and quasars in \citep{Yamashita} who used the HSC-SSP and FIRST data.
The HSC--SSP is an on-going optical imaging survey with five broadband filters ($g$-, $r$-, $i$-, $z$-, and $y$-band) and four narrowband filters \citep[see][]{Aihara_18a,Bosch,Coupon,Huang}.
This survey consists of three layers: Wide, Deep, and UltraDeep, and this work uses S16A Wide-layer data\footnote{The S16A data (Wide, Deep, and UltraDeep) will be available in 2019 as a public data release 2. Although \cite{Yamashita} used UltraDeep data in addition to Wide data, this work focuses only on Wide data.}obtained from 2014 March to 2016 January providing a forced photometry of $g$-, $r$-, $i$-, $z$-, and $y$-band with a 5$\sigma$ limiting magnitude of 26.8, 26.4, 26.4, 25.5, and  24.7, respectively \citep{Aihara_18b}.
The HSC--SSP Wide-layer covers six fields (XMM-LSS, GAMA09H, WIDE12H, GAMA15H, HECTOMAP, and VVDS; see Table 1 in \citealt{Yamashita} for detailed coordinates of each field).
The typical seeing is about 0$\arcsec$.6 in the $i$-band and the astrometric uncertainty is about 40 mas in rms.
Taking into account the photometric and astrometric flags, \cite{Yamashita} eventually extracted 23,795,523 HSC objects in the 154 deg$^2$ for the cross-matching with FIRST \cite[see Section 2.1 in][for more detail]{Yamashita}.

The FIRST project completed radio imaging survey at 1.4 GHz with a spatial resolution of 5$\arcsec$.4 \citep{Becker,White} covering 10,575 deg$^2$ that is completely overlapping with the survey footprint of the HSC-SSP Wide-layer, and the final release catalog of FIRST \citep{Helfand} is publicly available.
Before cross-matching with the HSC, \cite{Yamashita} made a flux-limited FIRST sample with flux density at 1.4 GHz greater than 1.0 mJy.
Taking into account a flag that tells a source is a spurious detection near a bright source, \cite{Yamashita} eventually extracted 7072 FIRST objects in the 154 deg$^2$ for the cross-matching with the HSC \citep[see Section 2.2 in][for more detail]{Yamashita}.
By cross-matching the HSC S16A Wide-layer catalog and FIRST final data release catalog with a search radius of 1$\arcsec$, 3579 objects (including RGs and radio-loud quasars) were selected \citep[see Section 3 in][for more detail]{Yamashita}.

Before compiling multi-wavelength data, we made a parent RG sample.
First, we removed 55 stellar objects (i.e., radio-loud quasars) based on optical morphological information \cite[see ][]{Yamashita}.
For 3579 -- 55 = 3524 RGs, we then narrowed down the sample to 2118 objects in three fields with a total area of $\sim 94.7$ deg$^{2}$ (GAMA09H, WIDE12H, and GAMA15H) where multi-wavelength data are available.
We then removed 175 objects that are not covered by FIR observation (see Section \ref{FIR}), which yielded 1943 RGs.
The sky distribution of those 1943 RGs is shown in Figure \ref{region}.
For those objects, we then complied the multi-wavelength data from $u$-band, near-IR (NIR), MIR, FIR, and radio data, as well as spectroscopic or photometric redshift.
After removing 897 objects with photometric data less than 10, and unreliable photometric redshift and/or photometric redshift greater than 1.7 (see Section \ref{CM}), we finally selected 1943 - 897 = 1056 objects (hereafter ``HSC--FIRST RGs'') with multi-wavelength data and reliable redshift in this work.

\subsubsection{$u$-band data}
The $u$-band data were taken from the Kilo-Degree Survey \citep[KiDS: ][]{de_Jong_13} that is an ESO public survey carried out with the VLT Survey Telescope (VST) and OmegaCAM camera \citep{Kuijken}.
We used the Data Release (DR) 3 \citep{de_Jong_17} that consists of 48,736,590 sources with a limiting magnitude of 24.3 mag (5$\sigma$ in a 2\arcsec aperture) in $u$-band.
The typical full width at half maximum (FWHM) of point spread function (PSF) for $u$-band detected point sources is about 1$\arcsec$\footnote{\url{http://kids.strw.leidenuniv.nl/DR3/catalog_table.php}.}.
Before the cross-matching, we extracted 42,252,797 sources with {\tt FLAG\_U} = 0 to ensure clean photometry in $u$-band \citep[see ][for more detail]{de_Jong_15,de_Jong_17}.

\subsubsection{Near-IR data}
We compiled NIR data from the VISTA Kilo-degree Infrared Galaxy Survey \citep[VIKING:][]{Arnaboldi} DR3\footnote{\url{http://eso.org/rm/api/v1/public/releaseDescriptions/107}} that includes 73,747,647 sources in $\sim$1000 deg$^{2}$ with NIR taken by the VISTA InfraRed Camera \citep[VIRCAM: ][]{Dalton}.
We used $J$-, $H$-, and $K$s-band with a median 10$\sigma$ (Vega) magnitude limit of 20.1, 19.0, and 18.6 mag, respectively.
Objects with a PSF FWHM of $<$ 1$\farcs$2 was observed in VIKING.
Before the cross-matching, we selected 63,028,265 objects with {\tt primary\_source} = 1 and ({\tt jpperrbits} $<$ 256 or {\tt hpperrbits} $<$ 256 or {\tt kspperrbits} $<$ 256) to ensure clean photometry for uniquely detected objects \citep[see also][]{Toba_15,Noboriguchi}.

\subsubsection{Mid-IR data}
\label{ALLWISE}
The MIR data were taken from {\it Wide-field Infrared Survey Explorer} \citep[{\it WISE}:][]{Wright}. 
We utilized W1 (3.4 $\micron$), W2 (4.6 $\micron$), W3 (12 $\micron$), and W4 (22 $\micron$) data in ALLWISE \citep{Cutri} that consists of 747,634,026 sources.
The 5$\sigma$ detection limits\footnote{\url{http://wise2.ipac.caltech.edu/docs/release/allwise/expsup/sec2_3a.html}} in W1, W2, W3, and W4 band are approximately 0.054, 0.071, 0.73 and 5 mJy, respectively.
The angular resolutions in W1, W2, W3, and W4 band are 6$\farcs$1, 6$\farcs$4, 6$\farcs$5, and 12$\farcs$0, respectively.
We extracted 741,753,366 sources with ({\tt w1sat} = 0 and {\tt w1cc\_map }= 0) or ({\tt w2sat} = 0 and {\tt w2cc\_map }= 0) or ({\tt w3sat} = 0 and {\tt w3cc\_map} = 0), or ({\tt w4sat} = 0 and {\tt w4cc\_map} = 0) in the AllWISE catalog \citep{Cutri}, to have secure photometry at either band (see the Explanatory Supplement to the AllWISE Data Release Products\footnote{\url{http://wise2.ipac.caltech.edu/docs/release/allwise/expsup/index.html}}, for more detail).

\subsubsection{Far-IR data}
\label{FIR}
We also used FIR data that were provided by a project of {\it the Herschel Space Observatory} \citep{Pilbratt} Astrophysical Terahertz Large Area Survey \citep[H-ATLAS: ][]{Eales,Bourne}.
The data were taken with the Photoconductor Array Camera and Spectrometer \citep[PACS: ][]{Poglitsch} at 100 and
160 $\micron$ and with the Spectral and Photometric Imaging REceiver instrument \citep[SPIRE: ][]{Griffin} at 250,
350, and 500 $\micron$.
The typical PSF FWHMs of 100, 160, 250, 350 and 500 $\micron$ are 11$\farcs$4, 13$\farcs$7, 17$\farcs$8, 24$\farcs$0 ,and 35\arcsec.2, respectively.
We used H-ATLAS DR1 \citep{Valiante} containing 120,230 sources in the GAMA fields.
The 1$\sigma$ noise for source detection (that includes confusion and instrumental noise) is 44, 49, 7.4, 9.4, and 10.2 mJy at 100, 160, 250, 350, and 500 $\micron$, respectively \citep{Valiante}.

\subsubsection{Ancillary Radio data}
The radio data were taken from observations with the Giant Metrewave Radio Telescope \citep[GMRT: ][]{Swarup}.
We used continuum flux density at 150 MHz ($\sim$1.99 m) provided by the Tata Institute of Fundamental Research (TIFR) GMRT Sky Survey (TGSS) alternative data release \citep[ADR:][]{Intema} that includes 623,604 radio sources in 36,900 deg$^2$.
The median rms noise of sources is 3.5 mJy beam$^{-1}$ with a spatial resolution of about 25\arcsec.

\subsubsection{Cross identification of multi-band catalogs}
\label{CM}
We then cross-identified those catalogs (KiDS, VIKING, ALLWISE, H-ATLAS, and TGSS) with HSC--FIRST RGs\footnote{We always use R.A. and Decl. in the HSC catalog as coordinates of HSC--FIRST objects.}.
By using a search radius of 1$\arcsec$ for KiDS and VIKING, 3$\arcsec$ for ALLWISE, 10$\arcsec$ for H-ATLAS, and 20$\arcsec$ for TGSS, 1051 (54.1\%), 1564 (80.5\%), 1482 (76.3\%), 257 (13.2\%), and 471 (24.2\%) objects were cross-identified by KiDS, VIKING, ALLWISE, H-ATLAS, and TGSS, respectively.
We note that 3/1051 ($\sim$0.3\%) and 2/471 ($\sim$0.4\%) objects have two candidates of counterpart for VIKING and TGSS sources, respectively within the search radius.
We choose the nearest object as a counterpart for such case.
For cross-matching with other catalogs (KiDS, ALLWISE, and H-ATLAS), one-to-one identification was realized. 
The matches by chance coincidence are estimated by generating mock catalogs with random positions, in the same manner as \cite{Yamashita}.
We generated mock catalog of KiDS, VIKING, ALLWISE, H-ATLAS, and TGSS data where source position in each catalog is shifted from the original one to $\pm$1$\arcdeg$ or $\pm$2$\arcdeg$ along the R.A. direction \citep[see][for more detail]{Yamashita}.
We then cross-identified HSC-FIRST RGs with those mock catalogs with the exactly same search radii.
We found that the chance coincidence of cross-matching with KiDS, VIKING, ALLWISE, H-ATLAS, and TGSS catalog is about 5.0, 1.9, 3.4, 9.3, and 0.6\%, respectively.

We also compiled photometric and spectroscopic redshift.
For spectroscopic redshift, we utilized the SDSS DR12 \citep{Alam}, the Galaxy and Mass Assembly project (GAMA) DR2 \citep{Driver,Liske}, and WiggleZ Dark Energy Survey project DR1 \citep{Drinkwater}. 
For photometric redshift, we employed a custom-designed Bayesian photometric redshift code \citep[{\tt MIZUKI}:][]{Tanaka} to estimate the photometric redshift (photo-$z$) of HSC--FIRST objects in the same manner as \cite{Yamashita} in which we used $z_{\rm best}$ as a photometric redshift \citep[see also][]{Tanaka_18}.
In order to perform an accurate SED fitting, we preferentially used spectroscopic redshift.
For objects without spectroscopic redshift, we used their $z_{\rm best}$ if they have a reliable photometric redshift, i.e., $0 < z_{\rm best} \leq  1.7$\footnote{\cite{Yamashita} reported that the HSC-SSP photo-$z$ derived by {\tt MIZUKI} could be secure at $z < 1.7$ based on comparison with spectroscopic redshift in COSMOS field \citep[see Section 5.1.2 in][for more detail]{Yamashita}}, $\sigma_{z_{\rm best}}/z_{\rm best} \leq  0.1$, and reduced $\chi^2$ of $z_{\rm best}$ $\leq $ 5.0.
These criteria are optimized based on the comparison with spectroscopic redshift for WERGS sample in \cite{Yamashita} \citep[see also][]{Tanaka_18}.
However, the influence of the above criteria on physical quantities derived from the SED fitting is still unclear, which will be discussed in Section \ref{inf_z}.
In addition to the above redshift (quality) cut, we extracted objects with 3$\sigma$ detection in at least 10 photometric bands among 20 photometric data ($u$, $g$, $r$, $i$, $z$, $y$, $J$, $H$, $K_{\rm s}$-band, and 3.4, 4.6, 12, 22, 100, 160, 250,  350, and 500 $\micron$, and 150 and 1400 MHz) to avoid an overfitting for our SED fitting method (see Section \ref{CIGALE}).
Consequently, 1056 HSC--FIRST RGs with multi-band photometry and reliable redshift were left (see Figure \ref{SS}).
Among 1056 objects, the redshifts of 224, 44, and 3 objects were taken from the SDSS DR12, GAMA DR2, and WiggleZ DR1, respectively while the redshifts of the remaining 785 objects were taken from {\tt MIZUKI}.
The HSC-FIRST RG catalog that includes basic information such as redshift and multi-band photometry is accessible through an online service.
Format and column descriptions of the catalog are summarized in Table \ref{catalog}.

\subsection{SED modeling with {\tt CIGALE}}
\label{CIGALE}

\begin{table}
\renewcommand{\thetable}{\arabic{table}}
\centering
\caption{Parameter ranges used in the SED fitting with {\tt CIGALE}.} 
\label{T1}
\begin{tabular}{lc}
\tablewidth{0pt}
\hline
\hline
Paramerer & Value\\
\hline
\multicolumn2c{Double exp. SFH}\\
\hline
$\tau_{\rm main}$ [Myr] & 1000, 3000, 4000, 6000 \\
$\tau_{\rm burst}$ [Myr] & 3, 5, 8, 15, 80 \\
$f_{\rm burst}$ & 0.001, 0.1, 0.3 \\
age [Myr] & 1000, 4000, 6000, 8000, 10000 \\
\hline
\multicolumn2c{SSP \citep{Bruzual}}\\
\hline
IMF & \cite{Chabrier} \\
Metallicity & 0.02 \\
\hline
\multicolumn2c{Dust attenuation \citep{Calzetti}}\\
\hline
& 0.01, 0.1, 0.15, 0.2, 0.25, \\
$E(B-V)_*$ &  0.3, 0.35, 0.4, 0.45, 0.5,\\
&  0.55, 0.6, 0.8, 1.0 \\
\hline
\multicolumn2c{AGN emission \citep{Fritz}}\\
\hline
$R_{\rm max}/R_{\rm min}$ &  60  \\
$\tau_{\rm 9.7}$ & 6.0 \\
$\beta$ & -0.50 \\
$\gamma$ & 0.0 \\
$\theta$ & 100.0 \\  
$\psi$ & 0.001, 60.100, 89.990 \\
$f_{\rm AGN}$ & 0.1, 0.5, 0.9\\
\hline
\multicolumn2c{Dust emission \citep{Dale}}\\
\hline
IR power-law slope ($\alpha_{\rm dust}$) & 0.0625, 0.2500, 1.0000, 2.0000 \\
\hline
\multicolumn2c{Radio synchrotron emission}\\
\hline
$L_{\rm FIR}/L_{\rm radio}$ coefficient ($q_{\rm IR}$) & 00.01, 0.1, 0.3, 0.5, 1.0, 2.5 \\
spectral index ($\alpha_{\rm radio}) $ & 0.5, 0.7, 0.9 1.1, 1.3 \\
\hline
\end{tabular}
\end{table}

We here employed {\tt CIGALE}\footnote{\url{https://cigale.lam.fr/2018/11/07/version-2018-0/}} \citep[Code Investigating GALaxy Emission:][]{Burgarella,Noll,Boquien} in order to perform a detailed SED modeling in a self-consistent framework with considering the energy balance between the UV/optical and IR.
In this code, we are able to handle many parameters such as star formation history (SFH), single stellar population (SSP), attenuation law, AGN emission, dust emission, and radio synchrotron emission. 

We assumed a SFH of two exponential decreasing SFR with different e-folding times \citep{Ciesla_15,Ciesla_16}.
We adopted the stellar templates provided from \cite{Bruzual} assuming the initial mass function (IMF) in \cite{Chabrier}, and the standard default nebular emission model included in {\tt CIGALE} \citep[see ][]{Inoue}.
Dust attenuation is modeled by using the \cite{Calzetti} law with color excess ($E(B-V)_*$).
We note that even if we employ the dust attenuation law of the Small Magellanic Cloud (SMC) that would be applicable to dusty starburst galaxies, resultant physical properties are consistent with what we present in this work within error.
The reprocessed IR emission of dust absorbed from UV/optical stellar emission is modeled assuming dust templates of \cite{Dale}.
For AGN emission, we also utilized models provided in \cite{Fritz} where we fixed some parameters that determines the density distribution of the dust within the torus to avoid a degeneracy of AGN templates in the same manner as \cite{Ciesla_15}.
We parameterized the $\psi$ parameter (an angle between the AGN axis and the line of sight) that corresponding to a viewing angle of the tours.
We also parameterize AGN fraction ($f_{\rm AGN}$) that is the contribution of IR luminosity from AGN to the total IR luminosity \citep{Ciesla_15}.
For radio synchrotron emission from either SFG or AGN, we parameterized a correlation coefficient between FIR and radio luminosity ($q_{\rm IR}$) and  the slope of power-law synchrotron emission ($\alpha_{\rm radio}$) (but see Sections \ref{Rindex} and \ref{S_qir}).
We define $\alpha_{\rm radio}$ from the measured radio flux density at observed-frame frequencies at 150 MHz and 1.4 GHz, assuming a power-law radio spectrum of $f_\nu \propto \nu^{-\alpha_{\rm radio}}$;
\begin{equation}
\label{index}
\alpha_{\rm radio} = \frac{\log \,(F_{150 \,{\rm MHz}} / F_{1.4 \,{\rm GHz}} )}{\log \,(\nu_{1.4\,{\rm GHz}} / \nu_{150\,{\rm MHz}})}
\end{equation}
This synchrotron emission is cut-off at 100 $\micron$ that is a default value adopted in {\tt CIGALE} that would be optimized for normal star-forming galaxies.
However, the synchrotron emission may contribute to fluxes/luminosities even at $< 100$ $\micron$ especially for radio-loud AGNs \citep[e.g.,][]{Mason,Privon,Falkendal,Rakshit}.
In this work, we choose 30 $\micron$ as a cutoff wavelength of the synchrotron emission with a single power-law, in the same manner as \cite{Lyu} \citep[see also][]{Pe'er}.
We have confirmed that the choice of cutoff wavelength does not significantly affect the following results.
Table \ref{T1} lists the detailed parameter ranges adopted in the SED fitting \citep[see also][]{Matsuoka,Chen,Toba}.
In addition to the energy balance between UV/optical and IR part, {\tt CIGALE} takes into account the balance between IR and radio luminosity that is parameterized by $q_{\rm IR}$, which are eventually an essential framework in {\tt CIGALE}.

In order to find a best-fit SED and calculate physical properties and their uncertainties, {\tt CIGALE} employed an analysis module so-called {\tt pdf\_analysis}.
This module computes the likelihood (that corresponds to $\chi^{2}$) for all the possible combinations of parameters and generate the probability distribution function (PDF) for each parameter and each object.
But before computing the likelihood, the module scaled the models by a factor ($\alpha$) to obtain physically meaningful values (so-called extensive physical properties) such as stellar masses and IR luminosities, where $\alpha$ can be derived as follows;
\begin{equation}
\alpha = \frac{\sum_i \frac{f_i  m_i}{\sigma_i^2}}{\sum_i \frac{m_i^2}{\sigma_i^2}} + \frac{\sum_j \frac{f_j m_j}{\sigma_j^2}}{\sum_j \frac{m_j^2}{\sigma_j^2}},
\end{equation}
where $f_{i}$ and $m_{i}$ are the observed and model flux densities, $f_{j}$ and $m_{j}$ are the observed and model extensive physical properties, and $\sigma$ is the corresponding uncertainties \citep[see Equation 13 in][]{Boquien}.
Finally, {\tt pdf\_analysis} computes the probability-weighted mean and standard deviation that correspond to resultant value and its uncertainty for each parameter, in which $\alpha$ is considered as a free parameter.
This approach is fully valid as far as one compare models built from the same set of parameters \cite[see Section 4.3 in ][for full expiation of this module]{Boquien} \cite[see also][]{Salim_07}.

Under the parameter setting described in Table \ref{T1}, we fit the stellar, AGN, and SF components to at most 20 photometric points ($u$, $g$, $r$, $i$, $z$, $y$, $J$, $H$, $K_{\rm s}$-band, and 3.4, 4.6, 12, 22, 100, 160, 250,  350, and 500 $\micron$, and 150 and 1400 MHz) of 1056 HSC--FIRST RGs observed with KiDS, HSC, VIKING, ALLWISE, H-ATLAS, FIRST, and TGSS.
For optical data, we used {\tt MAG\_AUTO\_U} as a $u$-band data that is a default magnitude\footnote{\url{http://www.eso.org/rm/api/v1/public/releaseDescriptions/82}} while {\tt g/r/i/z/ycModel\_Mag} were used for $g$-, $r$-, $i$-, $z$-, and $y$-band data \citep{Bosch,Huang}.
For NIR data, we used \cite{Petrosian} magnitude (see the release note of the VIKING DR3). 
Each magnitude were corrected for Galactic foreground extinction following \cite{Schlegel}.
The VIKING catalog contains the Vega magnitude of each source, and we converted these to AB magnitude, using offset values $\Delta m$ ($m_{\rm AB} = m_{\rm Vega} + \Delta m$) for $J$, $H$, and $K_{\rm s}$-band of 0.916, 1.366, and 1.827, respectively\footnote{\url{http://casu.ast.cam.ac.uk/surveys-projects/vista/technical/filter-set}}.
For MIR and FIR data, {\tt w1-4mpro} were utilized to estimate MIR flux densities \citep{Wright,Toba_14} while {\tt F100/160/250/350/500BEST} were used for FIR flux densities \citep[see][]{Valiante}.
ALLWISE catalog contains the Vega magnitude of each source, and we converted these to AB magnitude, using $\Delta m$ for 3.4, 4.6, 12, and 22 $\micron$ of 2.699, 3.339, 5.174, and 6.620, respectively\footnote{\url{http://wise2.ipac.caltech.edu/docs/release/allsky/expsup/sec4_4h.html\#conv2ab}}.
It is known that flux densities at 250, 350, and 500 $\micron$ could be boosted especially for faint sources (so-called ``flux boosting'' or ``flux bias'') that is caused by a confusion noise and instrument noise. 
Hence we corrected this effect by using the correction term provided in Table 6 of \cite{Valiante}.
For radio data, {\tt FINT} and {\tt STOTAL} were used for flux densities at 1.4 GHz and 150 MHz, respectively \citep[see ][for more detail]{Helfand,Intema}.
We used flux density at a wavelength when signal-to-noise ratio (S/N) is grater than 3 at that wavelength.
If an object was undetected, we put 3$\sigma$ upper limits at those wavelengths\footnote{{\tt CIGALE} can handle SED fitting of photometric data with upper limit in which they employed the method presented by \cite{Sawicki}. This method computes $\chi^{2}$ by introducing the error function (see Equations 15 and 16 in \cite{Boquien}.}.
Although the photometry employed in each catalog is different, their flux densities are expected to trace the total flux densities.
Therefore, the influence of different photometry is likely to be small.
Nevertheless, it is worth investigating whether or not physical properties can actually be estimated in a reliable way given an uncertainty of each photometry, which will be discussed in Section \ref{s_mock}.

\section{Results} 
\label{result}
\subsection{Histogram of i-band magnitude and redshift}
\label{hist}

\begin{figure}[h]
\plotone{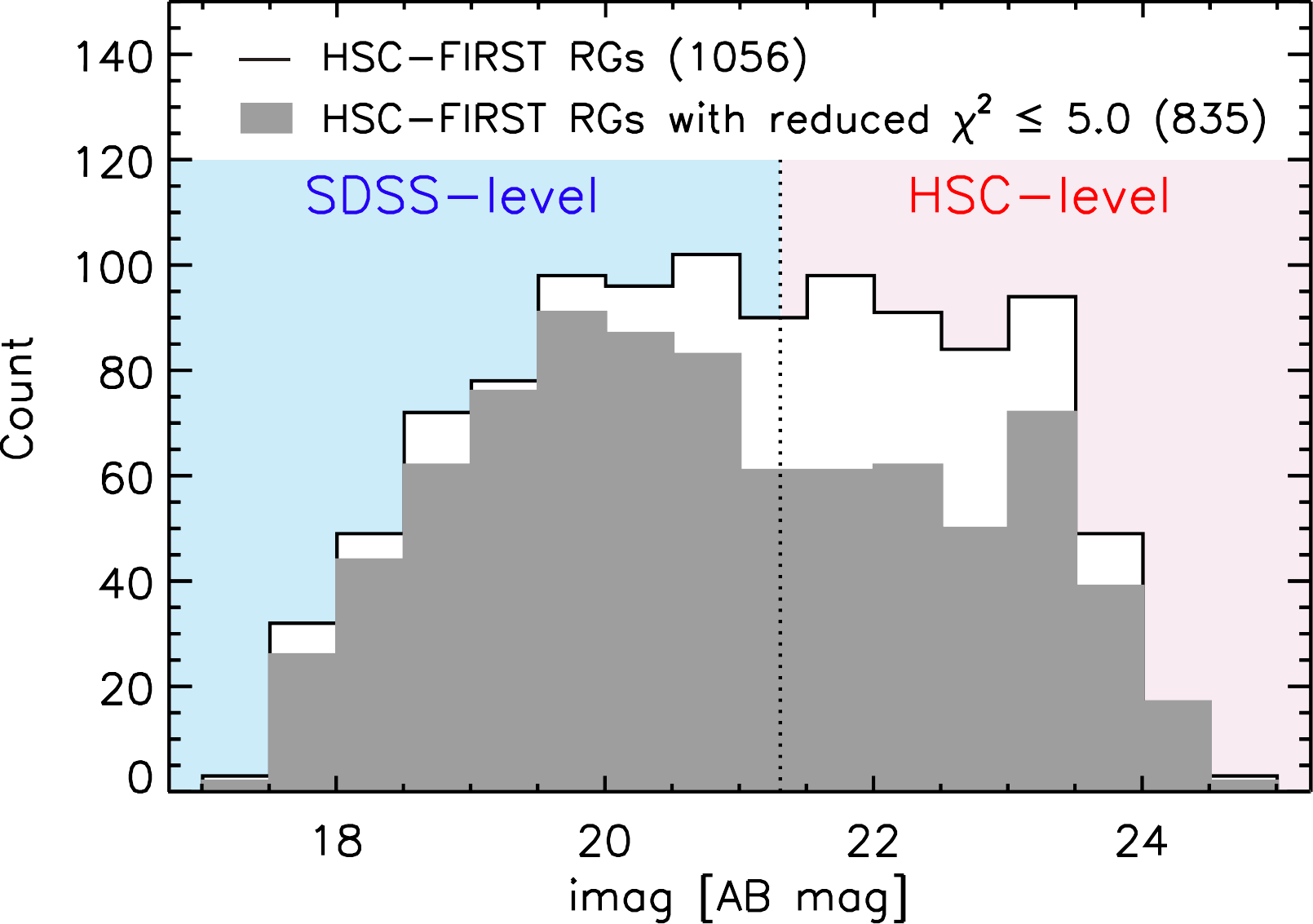}
\caption{Histogram of $i$-band magnitude of HSC--FIRST RGs (black line) and those with reduced $\chi^2$ of the SED fitting smaller than 5.0 (gray region), where $i$-band magnitude is corrected for the Galactic foreground extinction (see Section \ref{CIGALE}). The vertical dashed line is the threshold ($i$ = 21.3 mag) between 
SDSS-level and HSC-level RGs.}
\label{ihist}
\end{figure}

Figure \ref{ihist} shows a histogram of $i$-band magnitude for 1056 HSC--FIRST RGs.
Here, we define the ``SDSS-level objects'' and ``HSC-level objects'' based on the Galactic foreground extinction corrected $i$-band magnitude in the same manner as \cite{Yamashita}. 
We call objects with $i <$ 21.3 mag the SDSS-level objects as a reference of optically-bright RGs, while we call objects with $i \geq$ 21.3 mag the HSC-level objects as a reference of optically-faint RGs.
We found that 577 and 479 objects are classified as the SDSS-level and HSC-level objects, respectively, meaning that we have statistically robust sample of optically-faint RGs that are newly discovered by WERGS project \citep{Yamashita}.
\begin{figure}[h]
\plotone{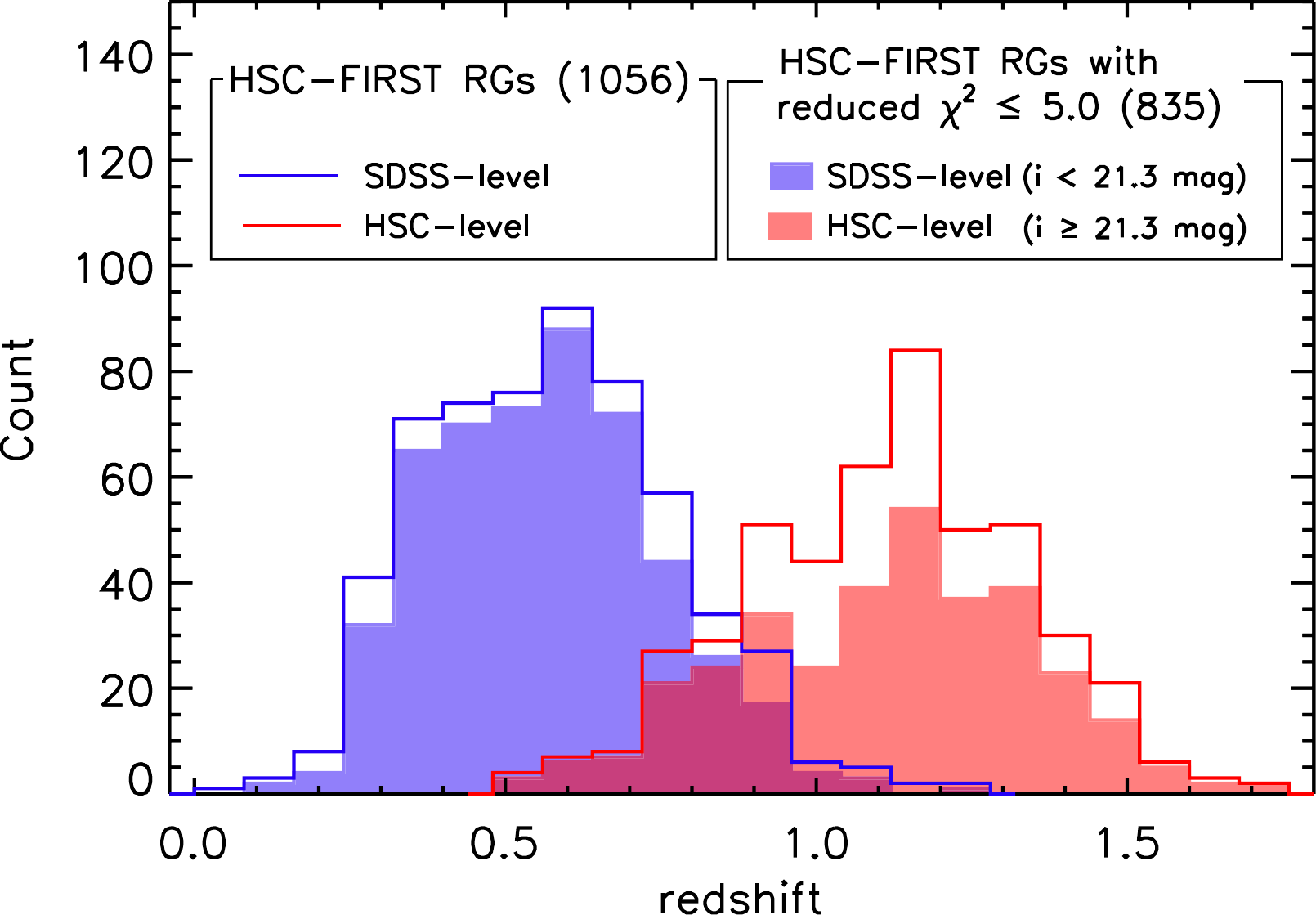}
\caption{Histogram of redshift of HSC--FIRST RGs (solid line) and those with reduced $\chi^2$ of the SED fitting smaller than 5.0 (shaded region). Red and blue line are the SDSS- and HSC-level objects in 1056 HSC--FIRST RGs. Red and blue regions are those in 835 subsample (see Section \ref{physical}).}
\label{redshift}
\end{figure}
Figure \ref{redshift} shows a histogram of redshift for 1056 HSC--FIRST RGs.
The mean values of redshift for the SDSS- and HSC- level objects are 0.57 and 1.10, respectively, meaning that HSC-level objects have larger redshift than SDSS-level objects, which is consistent with what \cite{Yamashita} reported.

\subsection{Result of SED fitting}

\begin{figure*}
\plotone{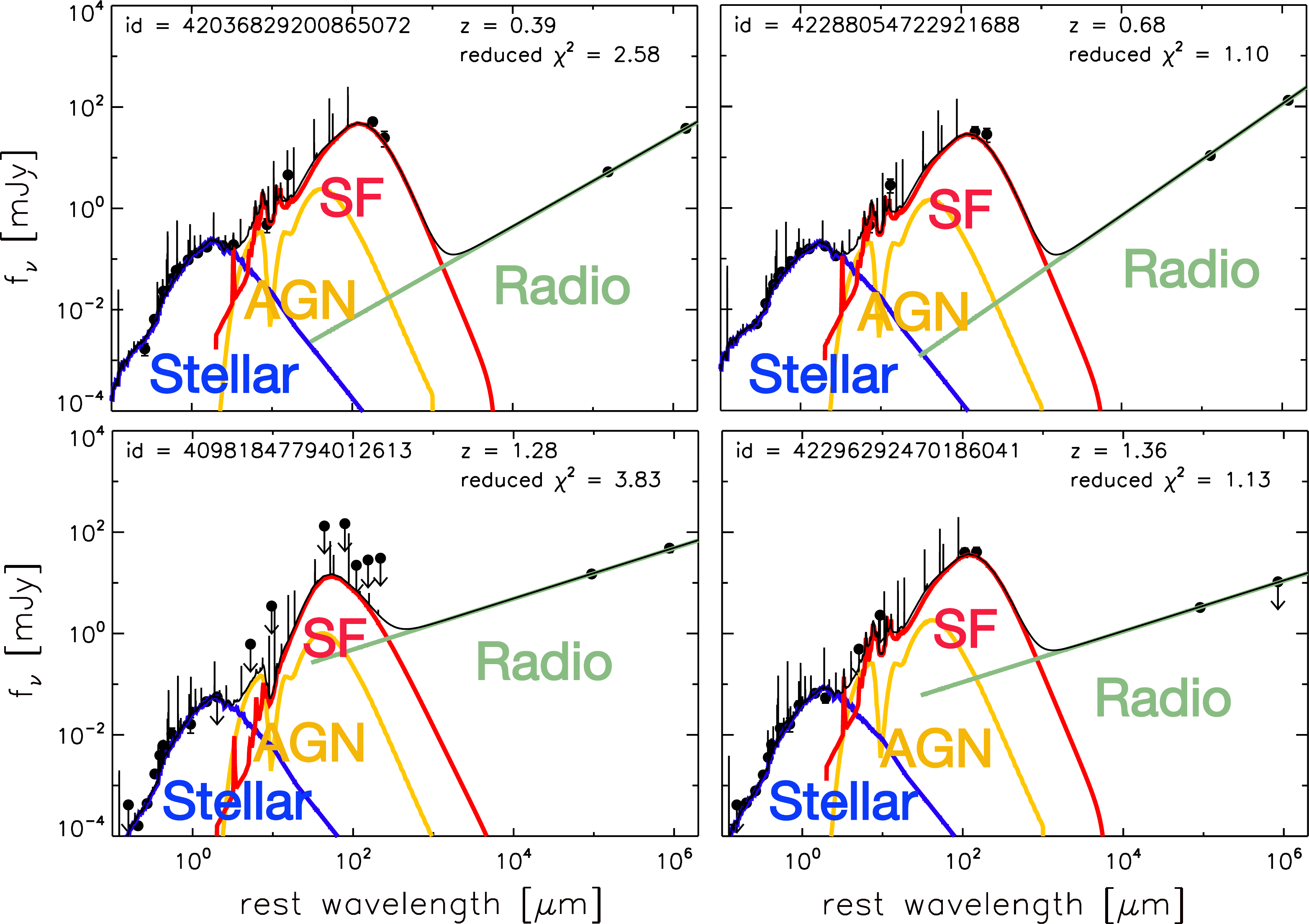}
\caption{Examples of the SED (flux density as a function of wavelength in rest-frame) and result of the SED fitting for our sample. The black points are photometric data where the down arrows mean 3$\sigma$ upper limit. The blue, yellow, red, and green lines show stellar, AGN, SF, and radio component, respectively. The black solid lines represent the resultant SEDs. We provide best-fit SEDs for all 1056 HSC--FIRST RGs with derived physical properties (see Tables \ref{catalog} and \ref{template}).}
\label{SED}
\end{figure*}

Figure \ref{SED} shows examples of the SED fitting with {\tt CIGALE}\footnote{Since {\tt CIGALE} assumed that the maximum wavelength for radio data was rest-frame 1 m, {\tt CIGALE} did not work for our dataset including TGSS (2 m) data for low-z objects. We modified {\tt CIGALE} code (radio.py) to solve this issue as suggested by Prof. Denis Burgarella through a private communication.}. 
We confirmed that 568/1056 ($\sim$54 \%) objects have reduced $\chi^2 \leq 3.0$ while 835/1056 ($\sim$79 \%) objects have reduced $\chi^2 \leq 5.0$, which means that the data are moderately well-fitted with the combination of the stellar, AGN, and SF components by {\tt CIGALE}.

We note that each quantity derived by the SED fitting would not be uniquely determined for some objects even if their reduced $\chi^2$ is good enough because there is a possibility of degeneracy among input parameters.
We checked the PDF of each quantity for randomly selected objects.
We confirmed that there is basically no prominent secondary peak of their PDFs, suggesting that the derived physical quantities are reliably determined.
The physical quantities such as stellar mass and SFR for 1056 HSC--FIRST RGs are also accessible through the online service (see Table \ref{catalog} for the catalog description).

\subsection{Radio and optical luminosity as a function of redshift}
\label{SzLM}

\begin{figure}
\plotone{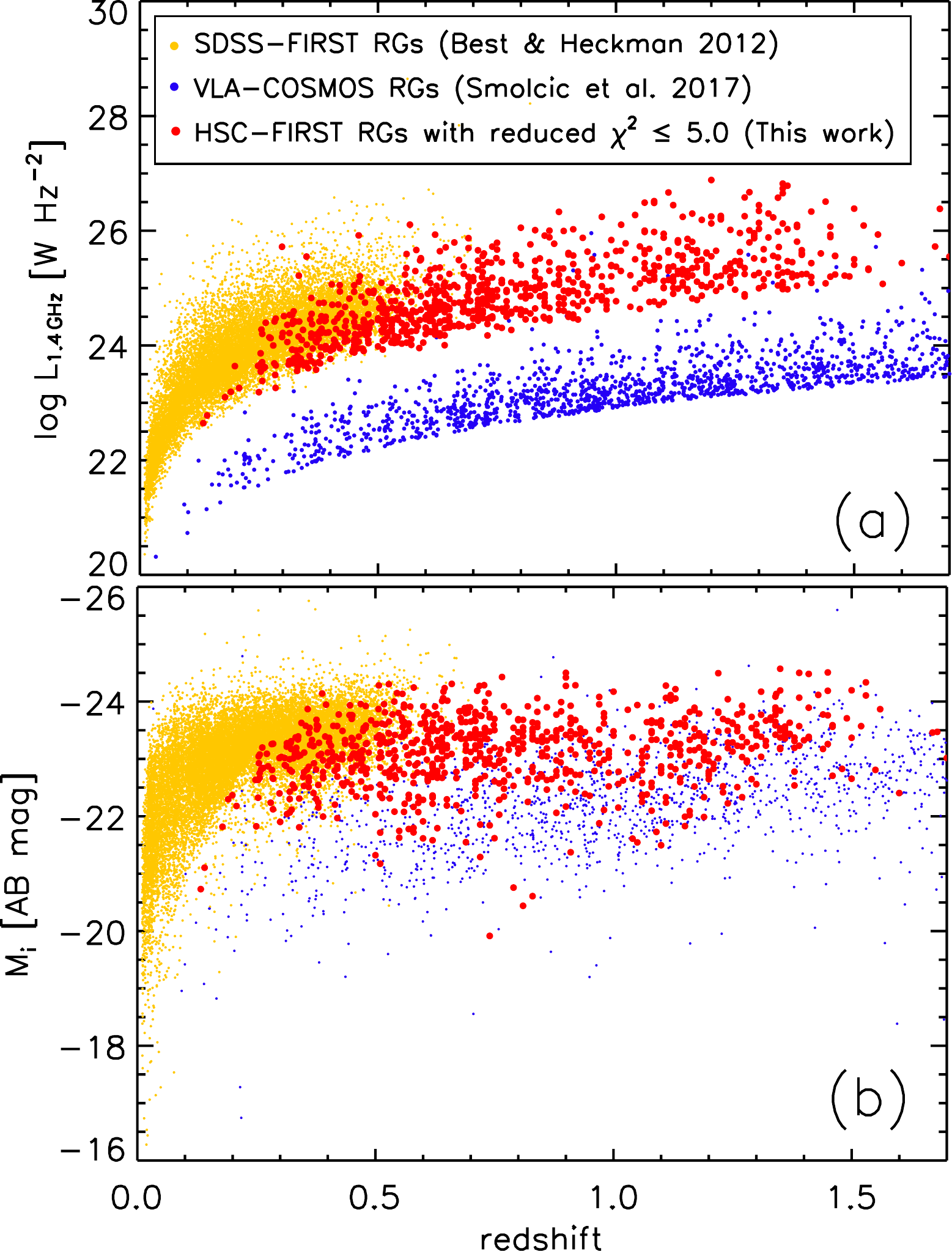}
\caption{(a) Rest-frame 1.4 GHz radio luminosity and (b) the absolute $i$-band magnitude at the rest frame as a function of redshift. Yellow and blue circles represent SDSS-detected RGs \citep{Best_12} and RGs discovered by VLA-COSMOS project \citep{Smolcic_17a}, respectively. Red circles represent HSC--FIRST RGs with reduced $\chi^{2} \leq$ 5.0.}
\label{zLM}
\end{figure}

Figure \ref{zLM}a shows the rest-frame 1.4 GHz radio luminosity ($L_{\rm 1.4\, GHz}$) of 835 HSC--FIRST RGs as a function of redshift.
In order to make sure the parameter space of our RGs with respect to previously discovered RGs, RGs selected with the SDSS \citep{Best_12} and RGs found by VLA-COSMOS 3 GHz large project \citep{Smolcic_17a,Smolcic_17b} are also plotted.
$L_{\rm 1.4\, GHz}$ in unit of W Hz$^{-1}$ is $k$-corrected luminosity at rest-frame 1.4 GHz that is derived by using formula;
\begin{equation}
\label{L14G}
L_{\rm 1.4 \,GHz} = \frac{4 \pi d_L^2 F_{\rm 1.4 \, GHz}}{(1+z)^{1- \alpha_{\rm radio}}},
\end{equation}
where $d_L$ is luminosity distance, $F_{\rm 1.4 \, GHz}$ is observed-frame flux density at 1.4 GHz, and $\alpha_{\rm radio}$ is radio spectral index we estimated in Equation \ref{index}.
We note that 190/835 objects have TGSS (150 MHz) data and thus their $\alpha_{\rm radio}$ are securely estimated.
If an object did not have $\alpha_{\rm radio}$ due to the non-detection of TGSS, we adopted a typical spectral index of RGs, $\alpha_{\rm radio}$ = 0.7 \citep[e.g.,][]{Condon_92} to estimate $L_{\rm 1.4\, GHz}$.
For radio sources selected either with the SDSS or VLA-COSMOS, we also used 0.7 as the spectral index to calculate $L_{\rm 1.4\, GHz}$ if the object did not have radio spectral index \citep[see e.g.,][]{Smolcic_17a}.
We confirmed that our RG sample distributes much higher redshift ($z > 0.5$) than SDSS-selected RGs while radio luminosity of our RGs sample is larger than that of VLA-COSMOS radio sources  with a median rms of 2.3 $\mu$Jy beam$^{-1}$.

Figure \ref{zLM}b shows the rest-frame $i$-band absolute magnitude ($M_{\rm i}$) as a function of redshift.
$M_{\rm i}$ of our RG sample was estimated based on the best-fit SED output by {\tt CIGALE}.
Since VLA--COSMOS catalog \citep{Smolcic_17a} does not contain $M_{\rm i}$, we used COSMOS2015 catalog \citep{Laigle_16} in which absolute magnitudes in optical and NIR bands were estimated based on the SED fitting.
For SDSS-selected RGs in \cite{Best_12}, we did not apply for any $k$-correction.
But their $M_{\rm i}$ can be approximately used for absolute magnitude at the rest-frame because they are low-$z$ objects.
We confirmed that our RG sample have intermediate value of $M_{\rm i}$ between SDSS-selected and VLA--COSMOS radio sources.

The discrepancy between our RG sample and VLA-COSMOS RG sample in $M_{\rm i}$ (Figure \ref{zLM}b) is much smaller than that in $L_{\rm 1.4\,GHz}$ (Figure \ref{zLM}a), suggesting that our RGs tend to trace higher radio-loudness sources, which is one of the advantages of WERGS project where even VLA-COSMOS might not be able to trace.
In summary, Figure \ref{zLM} reminds that our RG survey with HSC and FIRST explorers a new parameter space; relatively high-$z$ luminous radio galaxies, which is the advantage of this work.
We should keep in mind the above parameter space in the following discussions.

\subsection{WISE color-color diagram}
\label{WISE_2color}

\begin{figure*}
\plotone{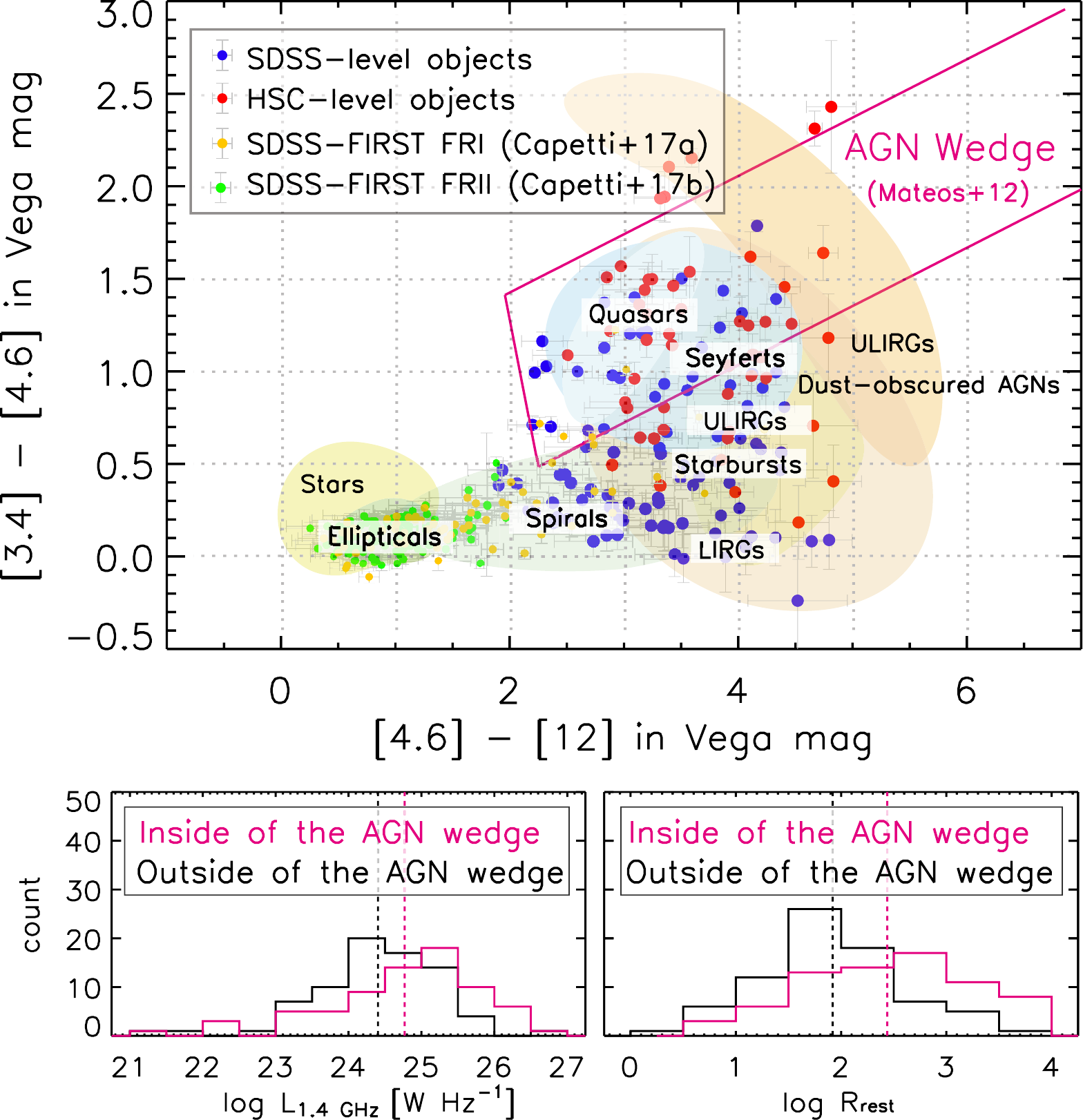}
\caption{(Top) {\it WISE} color-color diagram of 148 HSC--FIRST RGs with S/N $>$ 3 in 3.4, 4.6, and 12 $\micron$. Blue and red circles are SDSS- and HSC-level RGs, respectively. Yellow and green circles are SDSS-detected FIRST FR~I and FR~II RGs with $r <$ 17.8 mag, respectively that are obtained from \cite{Capetti_17a,Capetti_17b}. Regions with different color shading show typical MIR colors of different populations of objects \citep{Wright}. The solid lines illustrate the AGN selection wedge defined from \cite{Mateos_12,Mateos_13}. (Bottom) Histogram of rest-frame 1.4 GHz luminosity ($L_{\rm 1.4\,GHz}$) and rest-frame radio loudness ($R$) for objects inside (magenta) and outside (black) of the AGN wedge.The mean values are shown in dashed lines.}
\label{WISE}
\end{figure*}

Figure \ref{WISE} shows the {\it WISE} color-color diagram ([3.4] - [4.6] vs. [4.6] - [12]) for 148 HSC--FIRST RGs with S/N $>$ 3 in 3.4, 4.6, and 12 $\micron$ that were drawn from 1056 RG sample.
The anticipated MIR colors for various populations of objects are shown with different colors \citep{Wright}, which provides us a qualitative view of galaxies.
We found that the HSC-level objects tend to be redder than the SDSS-level objects in both colors of [3.4] - [4.6] and [4.6] - [12].
The majority of the SDSS-level objects is located at regions of spirals and LIRGs while the HSC-level objects are located at regions of Seyferts, Starburst galaxies, and ULIRGs.

About 49 \% of HSC-FIRST RGs with S/N $>$ 3 in 3.4, 4.6, and 12 $\micron$ are located within the AGN wedge defined by \cite{Mateos_12,Mateos_13}, who suggested a reliable MIR color selection criteria for AGN candidates based on the {\it WISE} and wide-angle Bright Ultrahard {\it XMM-Newton} survey \citep[BUXS:][]{Mateos_12}.
This means that roughly half of RG sample is outside of the wedge, which is in good agreement with previous works on radio-loud galaxies \citep{Gurkan_14,Banfield}, suggesting that the AGN selection based on the AGN wedge seems to be biased towards a subsample among the entire AGN population \citep[see also][]{Toba_14,Toba_15,Ichikawa}.

What makes the difference between objects in- and out-side of the AGN wedge?
One possibility is a difference of radio luminosity between them since radio luminosity is a good tracer of AGN power, as suggested by previous works \citep[e.g.,][]{Banfield,Singh_15,Singh_18}.
We checked this possibility for our sample, where we used the rest-frame radio luminosity at 1.4 GHz that is drawn from \cite{Yamashita} assuming a power-law radio spectrum of $f_\nu \propto \nu^{-0.7}$.

Figure \ref{WISE} shows the histogram of rest-frame 1.4 GHz luminosity, indicating a systematic difference in radio luminosity for objects in- and out-side of the AGN wedge.
The mean values of rest-frame 1.4 GHz luminosity for objects in- and out-side of the AGN wedge are $\log \,L_{\rm 1.4 \,GHz}$ $\sim$24.8 and $\sim$24.4 W Hz$^{-1}$, respectively, supporting the previous works.
An alternative indicator of AGN power is a radio loudness that is defined as flux ratio of rest-frame radio and optical band.
We used the radio loudness at rest-frame ($R_{\rm rest}$), a ratio of the rest-frame 1.4 GHz flux to the rest-frame
$g$-band flux as used in \cite{Yamashita}.
Figure \ref{WISE} also shows the histogram of $R_{\rm rest}$, indicating a systematic difference in $R_{\rm rest}$ for objects inside and outside of the AGN wedge.
The mean values of $R_{\rm rest}$ for objects in- and out-side of the AGN wedge are $\log \, R_{\rm rest}$ $\sim$ 2.4 and $\sim$1.9, respectively, indicating that objects with larger radio loudness tend to be located in the AGN wedge, as we expected.

We note that there are almost no objects at elliptical galaxies in the {\it WISE} color-color diagram (Figure \ref{WISE}), which is mainly interpreted as a selection bias of our HSC--FIRST RGs.
Since the saturation limit of the HSC for point sources at $r$-band and $i$-band are 17.8 and 18.4 mag, respectively \citep{Aihara_18b}, the HSC--FIRST RG sample does not contain those optically-bright objects.
In Figure \ref{WISE}, we also plot RGs with $r$-band magnitude smaller than 17.8 mag provided by \cite{Capetti_17a,Capetti_17b} who released \cite{Fanaroff} (FR) I and II RG catalogs\footnote{Since the catalogs do not contain {\it WISE} magnitudes, we cross-identified their {\it WISE} counterparts with a search radius of 3$\arcsec$ by ourselves.} selected with the SDSS and FIRST.
The redshift, optical absolute magnitude, and radio luminosity range of those RGs are $0.02 < z < 0.15$, $-23.7 < M_{\rm R} < -20.3$, and $23.3 < \log \,L_{\rm 1.4\,GHz} \,{\rm [W \,Hz^{-1}]} < 25.8$, respectively.
They show elliptical-like MIR colors, which means that optically too bright objects are located at region of elliptical galaxies.
In addition to the selection bias, there is a possibility that MIR colors of RGs would be different from normal elliptical galaxies.
\cite{Banfield} reported that [4.6] - [12] color of RGs selected from Radio Galaxy Zoo\footnote{\url{https://radio.galaxyzoo.org/}} sample shows significantly redder than that of typical elliptical galaxies.
This indicates that the dust emission of RGs may be enhanced compared with normal quiescent elliptical galaxies \citep[see also][]{Goulding,Gurkan_14}.
Indeed, \cite{Martini} reported that active elliptical galaxies tend to have a large dust mass compared with inactive elliptical galaxies, which supports the above hypotheses.

\subsection{Physical properties of HSC--FIRST radio galaxies}
\label{physical}
\begin{figure*}
\plotone{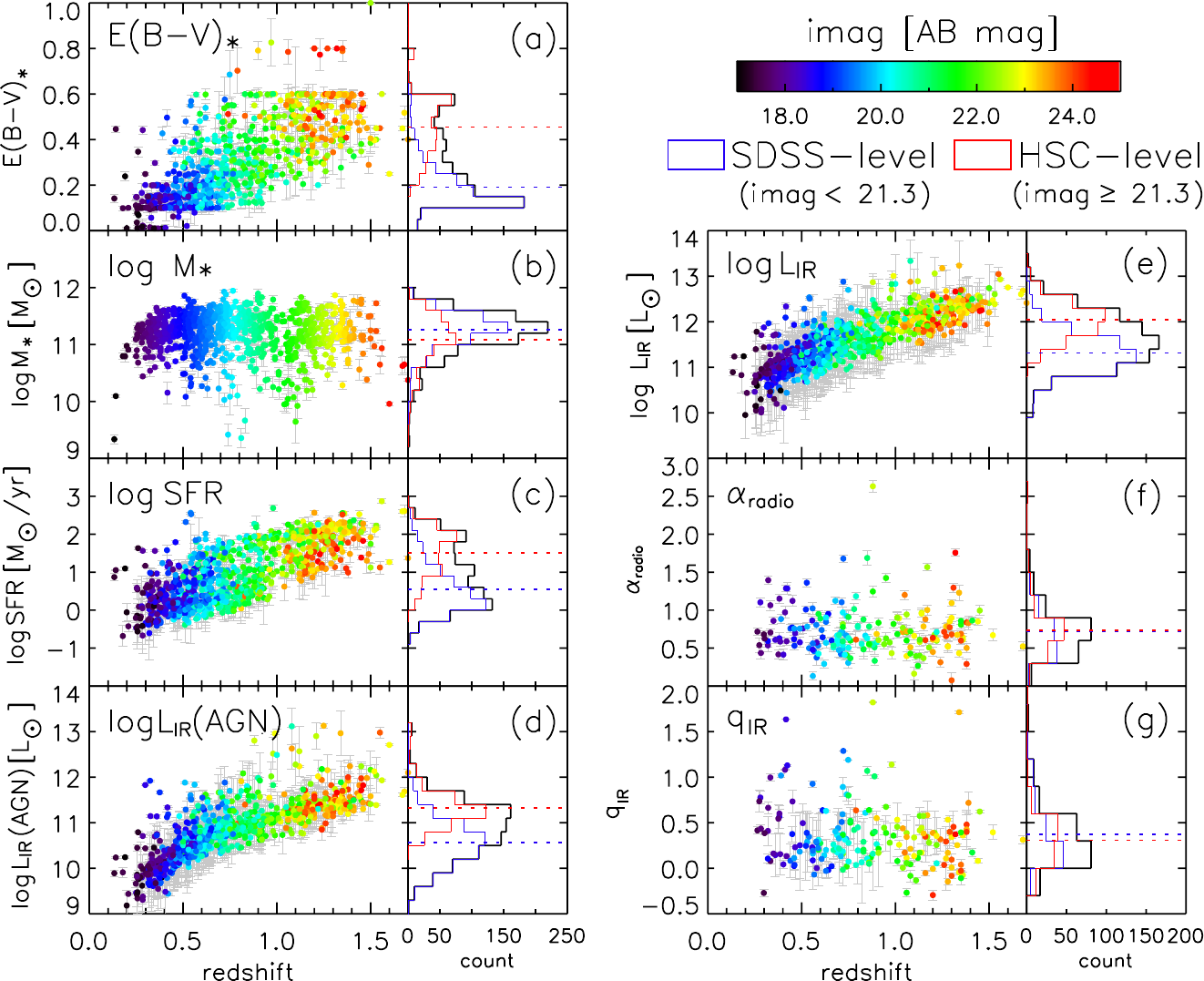}
\caption{(a) the color excess ($E(B-V)_{*}$), (b) stellar mass, (c) SFR, (d) IR luminosity contributed from AGNs, (e) total IR luminosity, (f) radio spectral index ($\alpha_{\rm radio}$), and (g) $q_{\rm IR}$ of HSC--FIRST RGs, as a function of redshift. The color code is $i$-band magnitude.
The histograms show the SDSS-level (blue), HSC-level (red), and total (black) objects. The dashed lines are mean values of each quantity for SDSS-level (blue) and HSC-level (red) objects. 835 RGs are plotted in panels (a) to (e) while 190 RGs with FIRST and TGSS data are plotted in panels (f) and (g).}
\label{phys}
\end{figure*}

We present the physical properties of HSC--FIRST RGs with being conducted a reliable SED fitting.
Hereafter, we will focus on a subsample of 835 HSC--FIRST RGs with reduced $\chi^2$ of the SED fitting smaller than  5.0.
In this work, we investigate the following quantities output directly from {\tt CIGALE}; (i) dust extinction, (ii) stellar mass, (iii) SFR, (iv) AGN luminosity, (v) IR luminosity, and those calculated by ourselves; (vi) radio spectral index, and (vii) $L_{\rm IR}/L_{\rm radio}$ coefficient ($q_{\rm IR}$), as a function of redshift, which are summarized in Figure \ref{phys}.
Among subsample, 501 and 334 objects are classified as the SDSS- and HSC-level objects with a mean redshift of 0.56 and 1.11, respectively (see Figures \ref{ihist} and \ref{redshift}).

\subsubsection{Dust extinction}
\label{S_EBV}

Figure \ref{phys}a shows color excess, $E (B-V)_{*}$, as a function of redshift, where $E (B-V)_{*}$ is an indicator of dust extinction of host galaxy.
We found that there is a clear correlation between redshift and $E(B-V)_{*}$; optically fainter RGs at high redshift are affected by larger dust extinction.
The mean values of $E(B-V)_{*}$ of the SDSS- and HSC- level objects are $\sim$0.19 and $\sim$0.45, respectively.
Indeed, 5 HSC--level objects with mean $E(B-V)_*$ of 0.45 satisfies a criterion of IR-bright dust-obscured galaxies with S/N $>$ 3 at 22 $\micron$ \citep[see e.g.,][]{Toba_15,Toba_16,Toba_17a,Toba_18,Noboriguchi}.

\subsubsection{Stellar mass}
\label{S_M}

Figure \ref{phys}b shows stellar mass as a function of redshift.
The stellar mass of our RG sample does not significantly depend on redshift, and thus the distributions of stellar masses for the SDSS- and HSC- level objects are similar.
However, the mean values of stellar mass of the SDSS- and HSC-level objects are $\log \,(M_{*}/M_{\sun})$ $\sim$11.26 and $\sim$11.08, respectively, indicating that the HSC-level RGs could tend to have less massive stellar mass compared with the SDSS-level ones.

\subsubsection{Star formation rate (SFR)}
\label{S_SFR}

Figure \ref{phys}c shows SFR as a function of redshift.
We found that the SFR increases with increasing redshift, and thus the HSC-level objects are systematically larger than those of the SDSS-level objects.
The mean values of SFR of the SDSS- and HSC-level objects are $\log$ SFR $\sim$ 0.55 and $\sim$1.51 $M_{\sun}$ yr$^{-1}$, respectively.
About one quarter of the HSC-level objects have SFR $>$ 100 $M_{\sun}$ yr$^{-1}$, which is consistent with what reported in WISE color-color diagram (Figure \ref{WISE}).

\subsubsection{AGN luminosity}
\label{Sec_AGN}

Figure \ref{phys}d shows IR luminosity contributed from AGN that is defined as $L_{\rm IR}\,$(AGN) = $f_{\rm AGN} \times L_{\rm IR}$ \citep{Ciesla_15} where $L_{\rm IR}$ is total IR luminosity (see Section \ref{Sec_IR}).
We found that the $L_{\rm IR}$\,(AGN) increases with increasing redshift, and thus the HSC-level objects seem to have systematically large AGN luminosity than the SDSS-level objects.
The mean values of $\log \, [L_{\rm IR}\,$(AGN)/$L_{\sun}]$ of the SDSS- and HSC-level objects are $\sim$ 10.56 and $\sim$11.32, respectively.

\subsubsection{IR luminosity}
\label{Sec_IR}

Figure \ref{phys}e shows IR luminosity as a function of redshift.
We can see a similar trend as AGN luminosity; IR luminosity increases with increasing redshift, and thus IR luminosities of the HSC-level objects are larger than those of the SDSS-level objects.
The mean values of $\log \, (L_{\rm IR}/L_{\sun})$ of the SDSS- and HSC-level objects are $\sim$11.31 and $\sim$12.04, respectively.
This is basically consistent with the fact that the majority of the SDSS- and HSC- objects is LIRGs and ULIRGs, respectively reported in Section \ref{WISE_2color}.

We note that since our RG sample may be affected by Malmquist bias as shown in Figure \ref{zLM}, the difference particularly in SFR, $L_{\rm IR}$ (AGN), and $L_{\rm IR}$ between SDSS- and HSC- level objects are basically due to the difference of their redshift distributions.
In other words, redshift dependence of $L_{\rm IR}$, $L_{\rm IR}$ (AGN), and SFR may be caused by sensitivity limit of IR bands.
On the other hand, it is natural that $M_*$ does not show a redshift dependence because the sensitivity of optical bands with HSC is much deeper than that of the IR bands.
If we compare SFR, $L_{\rm IR}$ (AGN), and $L_{\rm IR}$ of SDSS- and HSC-level objects at an overlapped redshift range (0.5 $< z <$ 1.0) (see Figure \ref{redshift}), the differences of mean values of SFR, $L_{\rm IR}$ (AGN), and $L_{\rm IR}$ are 0.31, 0.30, and 0.25 dex, respectively.
We also note that particularly SFR and AGN luminosity would also have an additional uncertainty probably due to a poor constraint of SED given a limited number of data points in MIR and FIR (see Section \ref{s_mock}).

\subsubsection{Radio spectral index}
\label{Rindex}

We present radio spectral index ($\alpha_{\rm radio}$) and luminosity ratio of IR and radio wavelength ($q_{\rm IR}$) in the following subsections.
Although our sample has always 1.4 GHz data, only one quarter of objects have 150 MHz data as reported in Section \ref{CM}.
This means that it is quite hard to determine the radio properties with {\tt CIGALE} for objects without counterparts of TGSS given a limited number of data points and input parameters.
Indeed, the radio spectral index and $q_{\rm IR}$ can be analytically derived by assuming a radio spectrum.
So, we focus on 190 HSC--FIRST RGs with both 1.4 GHz and 150 MHz flux densities in Subsections \ref{Rindex} and \ref{S_qir}.

We derive the radio spectral index ($\alpha_{\rm radio}$) based on Equation \ref{index}.
Figure \ref{phys}f shows radio spectral index as a function of redshift.
There is no clear correlation between $\alpha_{\rm radio}$ and redshift, which is consistent with previous works \citep{Blundell,Bornancini,Calistro}.
The mean value of $\alpha_{\rm radio}$ of 190 HSC--FIRST RGs is $\sim$0.73 that is consistent with what reported in \cite{de_Gasperin} who investigated radio spectral index over 80\% of the sky based on the NVSS and TGSS.
The mean values of $\alpha_{\rm radio}$ of the SDSS- and HSC-level objects are $\sim$0.72 and $\sim$0.74, respectively.
\cite{de_Gasperin} reported that the absolute value of radio spectral index increases with radio flux densities.
Since radio flux densities at 150 MHz and 1.4 GHz of the HSC-level objects are slightly larger than those of SDSS-level objects, the tiny difference of $\alpha_{\rm radio}$ between SDSS- and HSC-level objects could be explained as difference of their radio flux densities.

\subsubsection{$q_{\rm IR}$}
\label{S_qir}

The ratio of IR and radio luminosity ($q_{\rm IR}$) is defined as follows \citep[see also][]{Helou,Ivison};
\begin{equation}
\label{qir_eq}
q_{\rm IR} = \log \left( \frac{L_{\rm IR} / 3.75\times 10^{12}}{L_{\rm 1.4\, GHz}} \right),
\end{equation}
where $L_{\rm IR}$ is the total IR luminosity in unit of W derived from {\tt CIGALE}.
$3.75 \times 10^{11}$ is the frequency (Hz) corresponding to 80 $\micron$ that is used for making $q_{\rm IR}$ a dimensionless quantity.
$L_{\rm 1.4\, GHz}$ in unit of W Hz$^{-1}$ is $k$-corrected luminosity at rest-frame 1.4 GHz that is derived by Equation \ref{L14G}.

Figure \ref{phys}g shows $q_{\rm IR}$ as a function of redshift.
Although there is no clear dependence of $q_{\rm IR}$ on redshift, the mean value of 190 HSC--FIRST RGs is 0.34 that is significantly lower than that of pure SF galaxies whose $q_{\rm IR}$ is $\sim$ 2--3 \citep{Yun,Bell,Ivison}.
This is reasonable because it is known that radio-loud galaxies/AGNs with $\log \, L_{\rm radio} > 24$ W Hz$^{-1}$ tend to have significantly small $q_{\rm IR}$ with wide dispersion \citep{Sajina,Calistro,Williams}.
We will discuss this point later by using ``radio excess parameter'' in Section \ref{S_RE}.

The mean values of $q_{\rm IR}$ of SDSS- and HSC-level objects are $\sim$0.37 and $\sim$0.31, respectively.
\cite{Calistro} reported that $q_{\rm IR}$ could be decreased with increasing redshift while \cite{Read} reported that $q_{\rm IR}$ could also be decreased with increasing specific SFR (sSFR $\equiv$ SFR/$M_{*}$).
Since HSC-level objects are located at higher redshift and they have smaller stellar mass and higher SFR (i.e., higher sSFR) as mentioned in Sections \ref{S_M} and \ref{S_SFR}, the difference in $q_{\rm IR}$ between SDSS- and HSC-level objects may be explained by the difference of their redshift and sSFR.

\subsection{Composite spectrum}
 \begin{figure}
\plotone{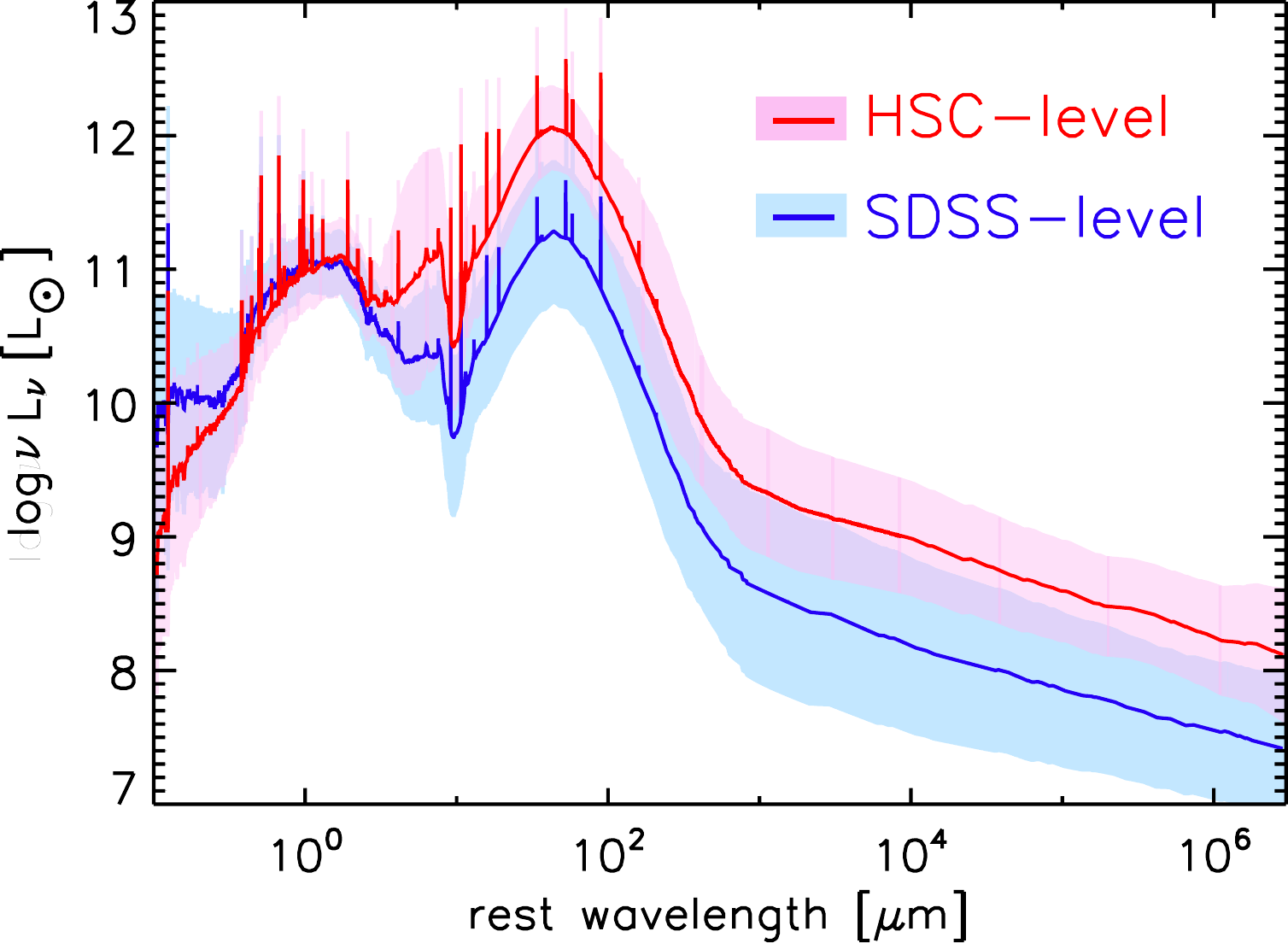}
\caption{Composite SEDs of SDSS- (blue) and HSC-level (red) RGs with reliable $\alpha_{\rm radio}$. Shared regions represent standard deviation of the median stacking SEDs. These SED templates are available in Table \ref{template}.}
\label{CompSED}
\end{figure}
Finally, we show a composite spectrum of the SDSS- and HSC-level objects in Figure \ref{CompSED}.
Here we performed the median stacking only for 190 HSC--FIRST RGs with reliable radio spectral index.
In optical to NIR regime, HSC-level objects are typically less luminous compared with SDSS-level objects, suggesting that HSC-level objects are more affected by dust extinction and their stellar masses are smaller than those of the SDSS-level objects as reported in Sections \ref{S_EBV} and \ref{S_M}.
Once wavelength is beyond 1 $\micron$, hot dust emission heated by AGNs and cold dust emission heated by SF will be dominant for HSC-level objects, indicating that HSC-level objects have a large AGN and SF luminosity (i.e, large IR luminosity and SFR) compared with SDSS-level objects as reported in Sections \ref{S_SFR}, \ref{Sec_AGN}, and \ref{Sec_IR}.
The best-fit SED template of each HSC--FIRST RG is available in Table \ref{template}.

\section{Discussion} 
\label{discussion}

\subsection{Selection bias}
\label{SB}
\begin{figure}
\plotone{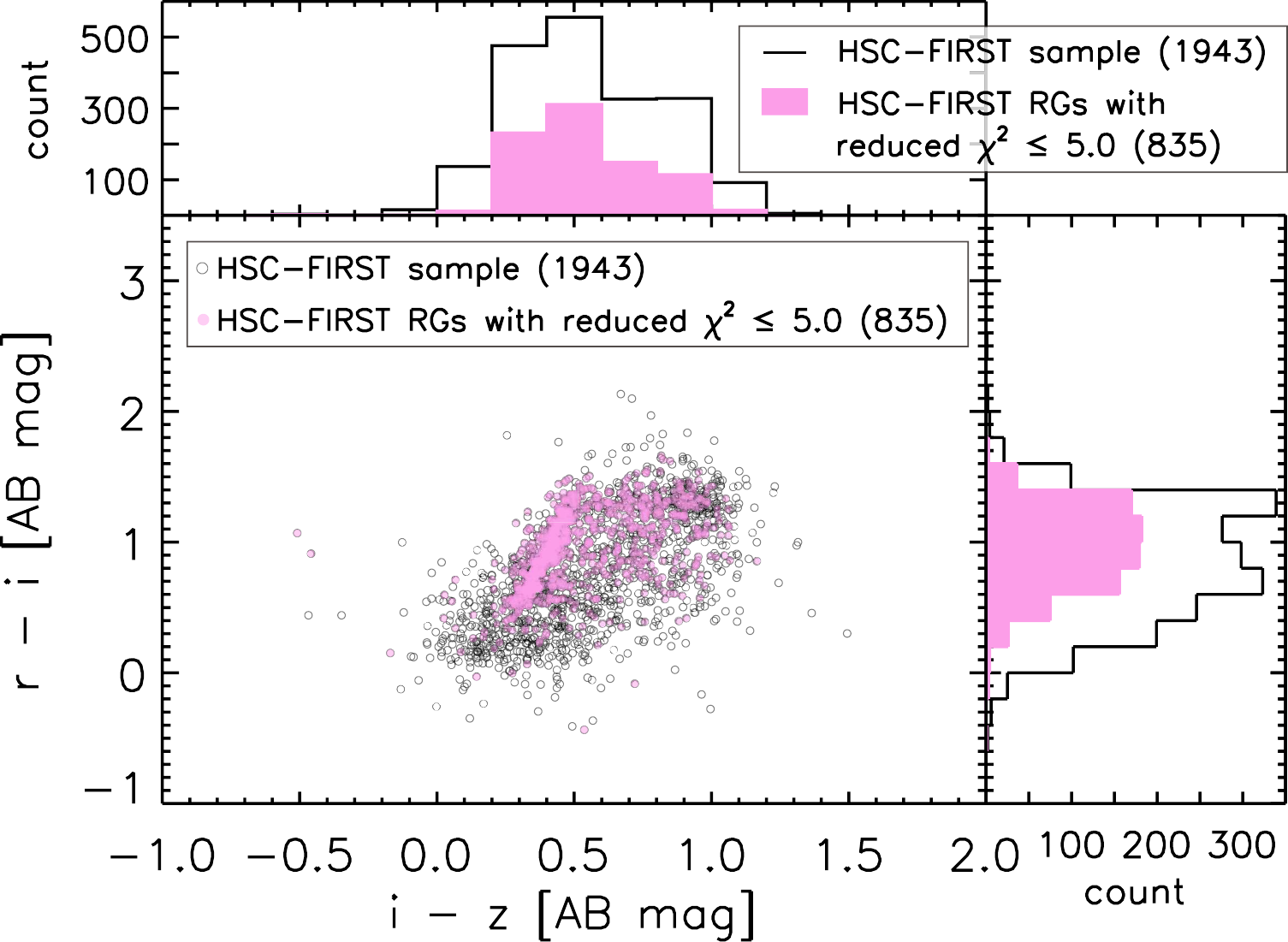}
\caption{Color-color diagram of $r - i$ and $i - z$. The 1943 HSC-FIRST RG sample, and 835 RGs whose physical properties are studied in this work, are shown in black and magenta circles, respectively. Histogram of each color is also shown with solid lines (an entire sample of 1943 objects) and magenta shaded regions (a subsample of 835 objects).}
\label{irz}
\end{figure}

As described in Sections \ref{DA} and \ref{result}, we selected 1056 objects with reliable redshift and reasonable redshift cut among 1943 RGs and eventually investigated physical properties for 835 RGs with SED fitting.
This means that 1943 - 835 = 1,108 ($\sim$57 \%) objects were excluded in this work, which would affect the results we presented the above.

In order to check whether or not we select a specific population among entire HSC--FIRST RG sample, we investigated their optical colors.
Figure \ref{irz} shows a color-color diagram of $r - i$ versus $i - z$ for entire sample of 1943 objects and subsample of 835 objects. 
Because HSC--FIRST RG sample requires all the detections of $r$, $i$, and $z$-band with S/N $>5$ \citep[see][]{Yamashita}, all objects in entire sample and subsample are plotted in this figure.
A two-sided K-S test does not rule out a hypothesis that the distribution of $i-z$ for the subsample of 835 RGs is same as that for the entire sample of 1943 RGs at $>$ 99.9\% significance, which is also supported by a Wilcoxon Rank-Sum test. 
On the other hand, those two tests find that two distributions of $r-i$ are statistically different.
This could suggest that physical quantities of subsample of 835 RGs may be (more or less) affected by selection bias that we should keep in mind in the following discussions.

\subsection{Possible uncertainties}
\label{PU}

We discuss the possible uncertainties of physical properties derived by {\tt CIGALE}.
We consider the following four things;  how (i) the uncertainty of photometric redshift and (ii) the difference in spatial resolution of each catalog affect the derived physical quantities, and comparison of resultant physical quantitates with (iii) spectroscopically derived ones and (iv) those derived from mock catalog.
We find that our RG sample is likely to have additional uncertainties especially for SFR and AGN luminosity.
However, it is hard to estimate the exact uncertainty for individual object because we infer the additional uncertainty based on a sort of Monte Carlo simulation.
Therefore, we do not include/propagate those possible uncertainties to the original ones output by {\tt CIGALE}, and focus on a statistical view of possible uncertainties.

\subsubsection{Uncertainty of photometric redshift}
\label{inf_z}
We selected 1056 HSC--FIRST RGs with reliable redshifts as described in Section \ref{DA}.
In particular, we allowed relative errors of photo-$z$ to be at most 10\%.
Here we discuss how the uncertainty of photo-$z$ affects the derived physical quantitates with SED fitting, by performing a following test.
First, we assumed a Gaussian distribution with a mean (a photo-$z$ of an object) and sigma (its photo-z error) for each object, and randomly choose one value among the distribution as an adopted redshift.
We then conducted the SED fitting with {\tt CIGALE} under the exact same parameter as what we used in this work for 785 objects whose redshifts are came from photo-$z$ with {\tt MIZUKI} (see Section \ref{CM}).

\begin{figure}
\plotone{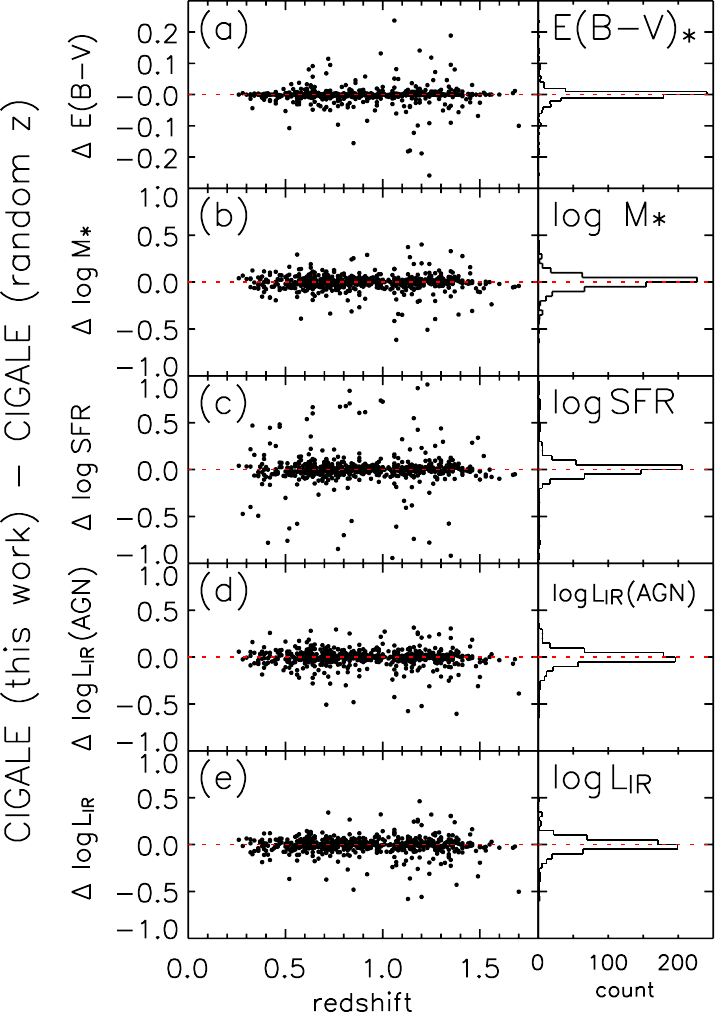}
\caption{The differences in $E(B-V)_{*}$, stellar mass, SFR, $L_{\rm IR}$ (AGN), and $L_{\rm IR}$ derived from {\tt CIGALE} in this work and those derived from {\tt CIGALE} with a random redshift assigned to each RG assuming a Gaussian probability function for the estimated photometric redshift. (a) $\Delta$$E(B-V)_{*}$, (b) $\Delta$$\log\, M_{*}$, (c) $\Delta$$\log$ SFR, (d)  $\Delta$$\log \,L_{\rm IR}$ (AGN), and (e) $\Delta \log\, L_{\rm IR}$ , as a function of redshift. The right panels show a histogram of each quantity. The red dotted lines are the $\Delta$ = 0.}
\label{z}
\end{figure}

Figure \ref{z} shows the differences in $E(B-V)_{*}$, $\log \,M_{*}$, $\log$ SFR, $\log \, L_{\rm IR}$ (AGN), and $\log \,L_{\rm IR}$ derived from {\tt CIGALE} in this work and those derived from {\tt CIGALE} with random redshift assuming a Gaussian for each object, as a function of redshift.
The mean values of each quantity are almost zero while the standard deviations of $\Delta$$E(B-V)_{*}$, $\Delta$$\log\, M_{*}$, $\Delta$$\log$ SFR, $\Delta$$\log \,L_{\rm IR}$ (AGN), and $\Delta \log\, L_{\rm IR}$ are 0.03, 0.09, 0.25, 0.10, and 0.10, respectively.
We found that $\Delta$$\log$ SFR is slightly larger than others due to a relatively large fraction of outliers, suggesting that SFR is most sensitive to uncertainty of photometric redshift.
We should keep in mind these possible uncertainty caused by photo-$z$ error.

\subsubsection{Influence of difference in spatial resolution of each catalog on physical quantities}
As described in Section \ref{DA}, we combined multi-wavelength catalogs with different spatial resolutions.
In particular, since the angular resolutions of {\it Herschel} and GMRT are relatively poor, we adopted 10$\arcsec$ and 20$\arcsec$ as a search radius to cross-identify with H-ATLAS and TGSS, respectively.
If there are multiple IR/radio sources within the search radii but H-ATLAS/TGSS could not resolve them, their FIR and radio (150 MHz) flux densities could be overestimated, which induces a systematic offset for physical quantities such as IR luminosity and radio spectral index that are derived by SED fitting \citep[see e.g.,][]{Pearson}.
This effect would be severe for fainter objects at high-$z$ Universe (i.e., HSC-level RGs).
If we could deblend those sources and re-measured FIR and radio flux densities for individual object, it would provide us (more or less) an accurate measurement of flux density although the deblending process may also have an uncertainty, which is beyond the scope of this paper. 
Therefore, we briefly discuss a possible influence of relatively large beam sizes of H-ATLAS and TGSS on derived physical quantities.

First, we check a possibility of overestimate of FIR flux densities in H-ATLAS by using ALLWISE catalog whose sensitivity and angular resolution are better than those of {\it Herschel} (see Section \ref{ALLWISE}).
We count all nearby {\it WISE} sources around an object with a search radius of 10$\arcsec$.
If more than one {\it WISE} sources are found around that object, those IR sources would contribute to FIR flux densities that are unresolved by {\it Herschel} and their FIR flux densities would be overestimated.
We confirm that 89/835 ($\sim11$ \%) objects have multiple {\it WISE} counterparts within 10$\arcsec$.
Here we test whether or not their IR luminosity has a systematically large value due to boost of their FIR flux densities.
\begin{figure}
\plotone{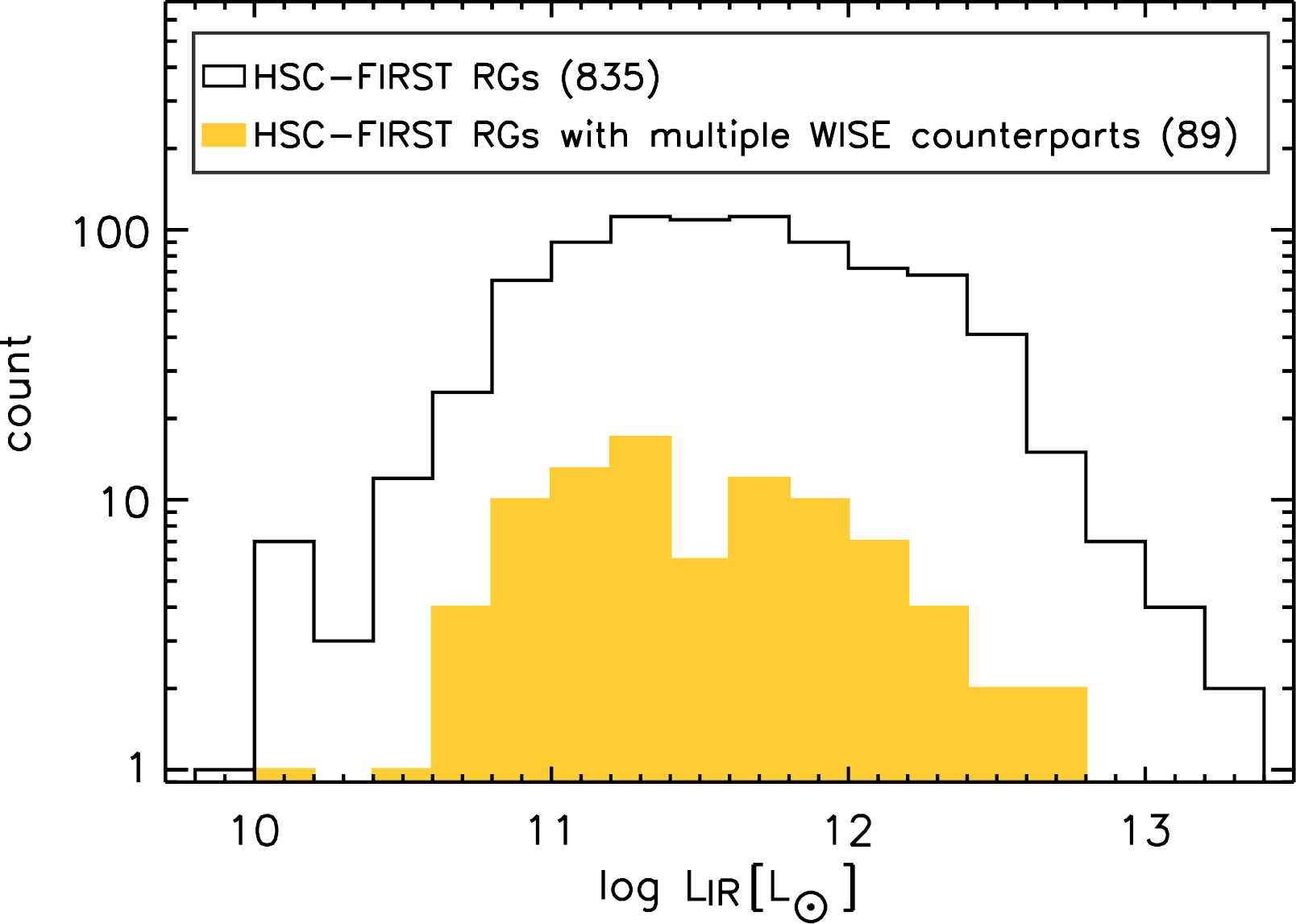}
\caption{The distribution of IR luminosity for HSC--FIRST RGs. Yellow shaded region corresponds to objects with multiple WISE counterparts.}
\label{LIR_multi}
\end{figure}
\begin{figure}
\plotone{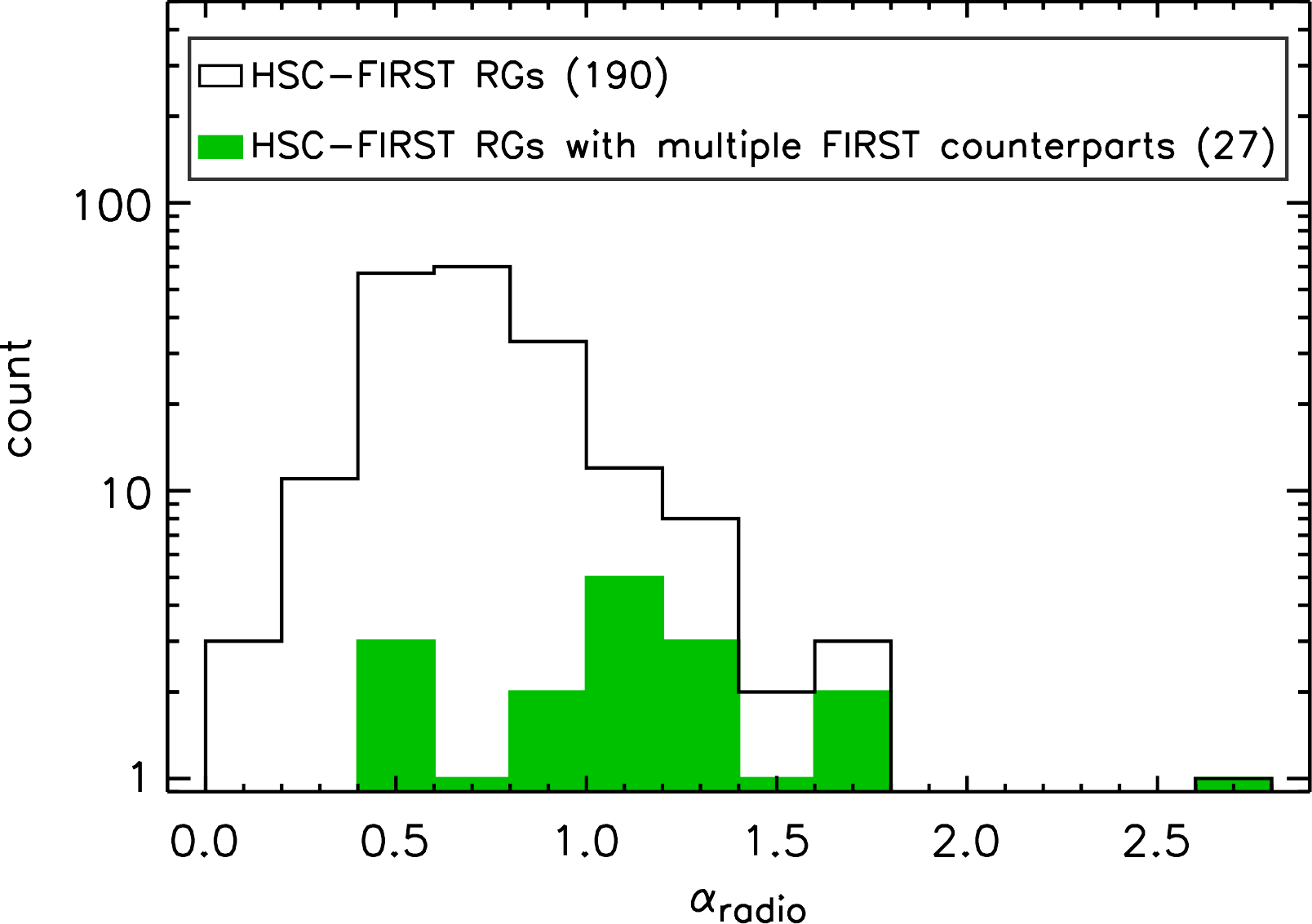}
\caption{The distribution of radio spectral index for HSC--FIRST RGs. Green shaded region corresponds to objects with multiple FIRST counterparts.}
\label{alpha_multi}
\end{figure}

Figure \ref{LIR_multi} shows the histogram of IR luminosity for 835 HSC-FIRST RGs and 89 objects with multiple {\it WISE} counterparts.
We find that there is no systematic difference between them.
The mean IR luminosity of 89 objects is $\log\, (L_{\rm IR}/L_{\sun}) \sim 11.48$ that is in good agreement with that of all HSC--FIRST RGs, suggesting that poor angular resolution of H-ATLAS does not significantly affect the measurement of FIR flux densities.

Next, we check a possibility of overestimate of radio flux density at 150 MHz in TGSS by using FIRST catalog whose sensitivity and angular resolution (6$\arcsec$) are better than those of GMRT.
We count all nearby FIRST sources around an object with a search radius of 20$\arcsec$, and confirm that 27/190 ($\sim14$ \%) objects have multiple FIRST counterparts within 20$\arcsec$.
Here we test whether or not their radio spectral index ($\alpha_{\rm radio}$) have systematically large value due to boost of their 150 MHz flux densities.
Figure \ref{alpha_multi} shows the histogram of radio spectral index for 190 HSC-FIRST RGs and 27 objects with multiple FIRST counterparts.
We find that 27 objects systematically have large $\alpha_{\rm radio}$.
Their mean $\alpha_{\rm radio}$ is 1.19 that is significantly larger than that of all HSC--FIRST RGs, suggesting that radio spectral indices of some RGs have a potential to be overestimated.
We note that radio morphology of some RGs looks different from optical/IR; for example, they have radio lobes in addition to radio core, which makes the cross-identification between optical and radio complicated.
We visually checked radio images to see how many RGs could have that kind of complex morphology.
We found that 48/835 ($\sim$5.7 \%) of our RGs sample would have such morphology.
Their mean $\alpha_{\rm radio}$ is 1.05 that is also larger than the typical value of HSC--FIRST RGs, suggesting that flux density at 150 MHz taken by TGSS with poor spatial resolution may measure even emission from lobes and thus their $\alpha_{\rm radio}$ may be overestimated.

\subsubsection{Comparison with spectroscopically derived quantities}

We derived $E(B-V)_{*}$, stellar mass, and SFR based on photometric data with SED fitting as presented in Sections \ref{DA} and \ref{result}.
Here, we check the consistency between those quantities derived based on {\tt CIGALE} and spectroscopic data.
We compiled the stellar masses from the SDSS DR12 {\tt stellarMassPCAWiscBC03} table that are derived using the method of \cite{YChen} with the SSP models of \cite{Bruzual}.
Since a default IMF adopted in {\tt stellarMassPCAWiscBC03} table is \cite{Kroupa}, we converted their Kroupa stellar masses to those with \cite{Chabrier} IMF by subtracting 0.05 dex from the logarithm of stellar masses, in the same manner as \cite{YChen}.
For $E(B-V)_{*}$, we utilized the SDSS DR12 {\tt emissionLinesPort} table in which objects are fitted using an adaptation of the publicly available Gas AND Absorption Line Fitting \citep[GANDALF;][]{Sarzi} and penalised PiXel Fitting \citep[pPXF;][]{Cappellari}.
Stellar population models for the continuum are come from \cite{Maraston} and \cite{Thomas}.
For SFR, we used an emission line-based SFR where we selected [O{\,\sc ii}] $\lambda\lambda$3726,3729 doublet that is known as a good indicator of SFR \citep[e.g.,][]{Kennicutt}.
We used a relation suggested by \cite{Kewley} to estimate [O{\,\sc ii}]-based SFR (SFR$_{\rm [OII]}$);

\begin{equation}
{\rm SFR}_{\rm [OII]} = (6.58 \pm 1.65) \times 10^{-42} L_{\rm [OII]}^{\rm cor},
\end{equation}
where $L_{\rm [OII]}^{\rm cor}$ is the extinction-corrected [O{\,\sc ii}] luminosity in units of erg s$^{-1}$ that is calculated using the following formula \citep[see][]{Calzetti_94,Dominguez};
\begin{equation}
L_{\rm [OII]}^{\rm cor} = L_{\rm [OII]}^{\rm obs} 10^{0.4 k_{\rm [OII]}\, E(B-V)_{\rm gas}},
\end{equation}
where $L_{\rm [OII]}^{\rm obs}$ is the observed [O{\,\sc ii}] luminosity, $k_{\rm [OII]}$ is the extinction value  at $\lambda = 3727$ \AA \,\,provided by \cite{Calzetti}, and $E(B - V)_{\rm gas}$ is the color excess estimated from emission lines.
The observed [O{\,\sc ii}] flux and $E(B - V)_{\rm gas}$ are tabulated in {\tt emissionLinesPort} table. 

\begin{figure}
\plotone{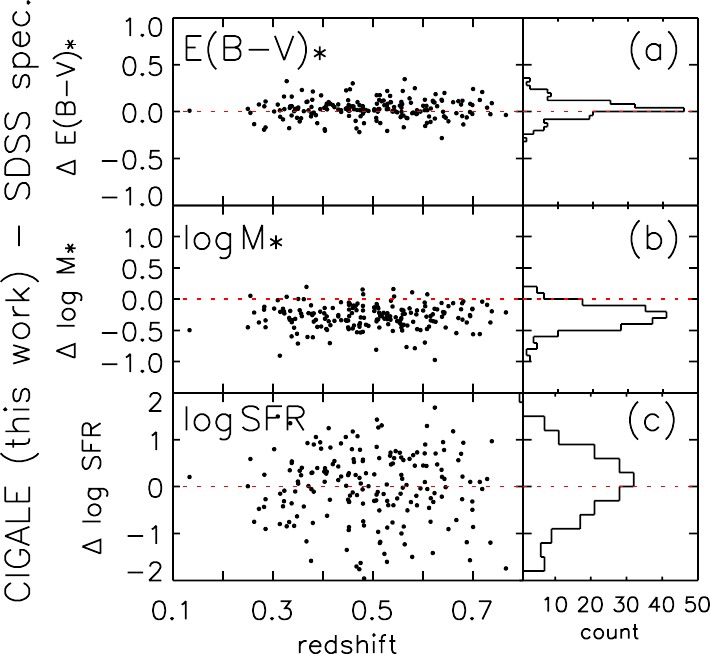}
\caption{The differences in $E(B-V)_{*}$, stellar mass, and SFR derived from {\tt CIGALE} and those derived from the SDSS DR12 spectroscopic data ({\tt stellarMassPCAWiscBC03} and {\tt emissionLinesPort} table). (a) $\Delta$$E(B-V)_{*}$, (b) $\Delta$$\log\, M_{*}$, and (c) $\Delta$$\log$ SFR, as a function of redshift. The right panels show a histogram of each quantity. The red dotted lines are the $\Delta$ = 0.}
\label{sdss}
\end{figure}

Figure \ref{sdss} shows the differences in $E(B-V)_{*}$, stellar mass, and SFR derived from {\tt CIGALE} and those derived from the SDSS spectroscopic data (i.e., {\tt stellarMassPCAWiscBC03} and {\tt emissionLinesPort} table).
We found that $E(B-V)_{*}$ derived from {\tt CIGALE} is slightly overestimated by 0,03 dex while $\log\, M_{*}$ derived from {\tt CIGALE} is significantly underestimated by 0.27 dex (see Figure \ref{sdss}ab).
However, this offset is consistent with what reported in \cite{YChen} who compared stellar masses derived from their method with principal component analysis (PCA) and those derived from the SDSS 5-band photometry.
They reported the PCA-based stellar mass shows a systematically positive offset.
We also note that assumed SFH in \cite{YChen} differs from that in this work, which would also induce a systematic difference of $E(B-V)_{*}$ and stellar mass.
The mean value of $\Delta$ $\log$ SFR is 0.06 that is negligibly small while its standard deviation is 0.77 that is very large as shown in Figure \ref{sdss}c.
Because a typical uncertainty of [OII]-based SFR is about 0.6 dex, whether or not the above large offset is significant is still unclear. 
An another possibility of the large dispersion of $\Delta$ $\log$ SFR may be a contamination of AGN extended
emission line region.
Recently, \cite{Maddox} reported that [O{\,\sc ii}] is not always a good indicator of SFR for AGNs when strong [Ne{\,\sc v}]$\lambda3426$ is present in the AGN spectrum.
Roughly a quarter of RG sample with SDSS spectra has prominent [Ne{\,\sc v}] lime with S/N $>$ 5.0, and thus their [O{\,\sc ii}]-based SFR would have a large uncertainty.
Nevertheless, we should keep in mind the possibility of those systematic uncertainness.
On the other hand, this test is only appreciable to SDSS-level objects ($z <$ 0.8) and thus we need to check whether or not the resultant quantities of HSC-level objects is reliable through an another way (see Section \ref{s_mock}).

\subsubsection{Comparison with physical quantities derived from mock catalog}
\label{s_mock}
Since {\tt CIGALE} has a procedure to asses whether or not physical properties can actually be estimated in a reliable way through the analysis of a mock catalog, we here discuss the influence of photometric uncertainty on the derived physical quantities.
In order to make the mock catalog, {\tt CIGALE} first uses the photometric data for each object based on the best-fit SED, and then modify each photometry by adding a value taken from a Gaussian distribution with the same standard deviation as the observation. 
This mock catalogue is then analyzed in the exact same way as the original observations \cite[see][for more detail]{Boquien}.

\begin{figure}
\plotone{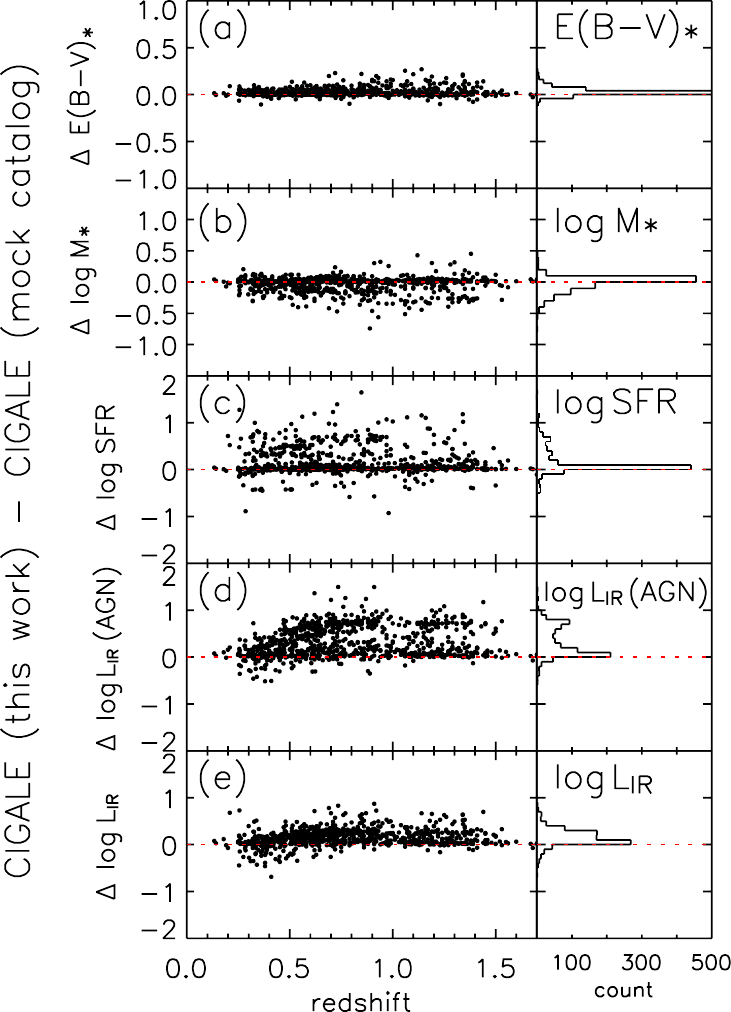}
\caption{The differences in $E(B-V)_{*}$, stellar mass, SFR, $L_{\rm IR}$ (AGN), and $L_{\rm IR}$ derived from {\tt CIGALE} in this work and those derived from mock catalog. (a) $\Delta$$E(B-V)_{*}$, (b) $\Delta$$\log\, M_{*}$, (c) $\Delta$$\log$ SFR, (d) $\Delta$$\log \,L_{\rm IR}$ (AGN), and (e) $\Delta \log\, L_{\rm IR}$, as a function of redshift. The right panels show a histogram of each quantity. The red dotted lines are the $\Delta$ = 0.}
\label{mock}
\end{figure}

Figure \ref{mock} shows the differences in $E(B-V)_{*}$, stellar mass, SFR, $L_{\rm IR}$ (AGN), and $L_{\rm IR}$ derived from {\tt CIGALE} in this work and those derived from mock catalog, as a function of redshift.
The mean values of $\Delta$$E(B-V)_{*}$, $\Delta$$\log\, M_{*}$, $\Delta$$\log$ SFR, $\Delta$$\log \,L_{\rm IR}$ (AGN), and $\Delta \log\, L_{\rm IR}$ are 0.03, -0.03, 0.14, 0.15, and 0.32, respectively.
In particular, we can see a secondary peak in $\Delta$$\log$ SFR, $\Delta$$\log \,L_{\rm IR}$ (AGN) regardless of redshift.
This suggests that SFR and AGN luminosity are sensitive to uncertainty of photometry, which may be a limitation of our SED fitting method given a limited number of data points in MIR and FIR.

\subsection{Stellar mass and SFR relation as a function of redshift}
\label{MSFR}
It is well known that stellar mass and SFR of galaxies are correlated, and the majority of galaxies follow a relation called the ``main sequence (MS)'' \citep[e.g.,][]{Brinchmann,Daddi,Elbaz}.
This relation is evolved toward high redshift \citep[e.g.,][]{Speagle,Lee,Tomczak}.
Galaxies undergoing active SF (so-called starburst galaxies) lie above the MS while those without active SF (so-called passive galaxies) lie below the MS.
The stellar mass and SFR are fundamental physical quantities of galaxies, and thus investigating the relation ($M_{*}-$SFR) provides us a clue of galaxy evolution.
Here we investigate the stellar mass and SFR relation for HSC--FIRST RGs to see if there is any difference between SDSS- and HSC-level RGs. 
Since stellar masses and SFRs of RGs depend on $i$-band magnitude and redshift (see Figure \ref{phys}bc), we check $M_{*}-$SFR for SDSS- and HSC-level RGs, as a function of redshift.

\begin{figure}[h]
\plotone{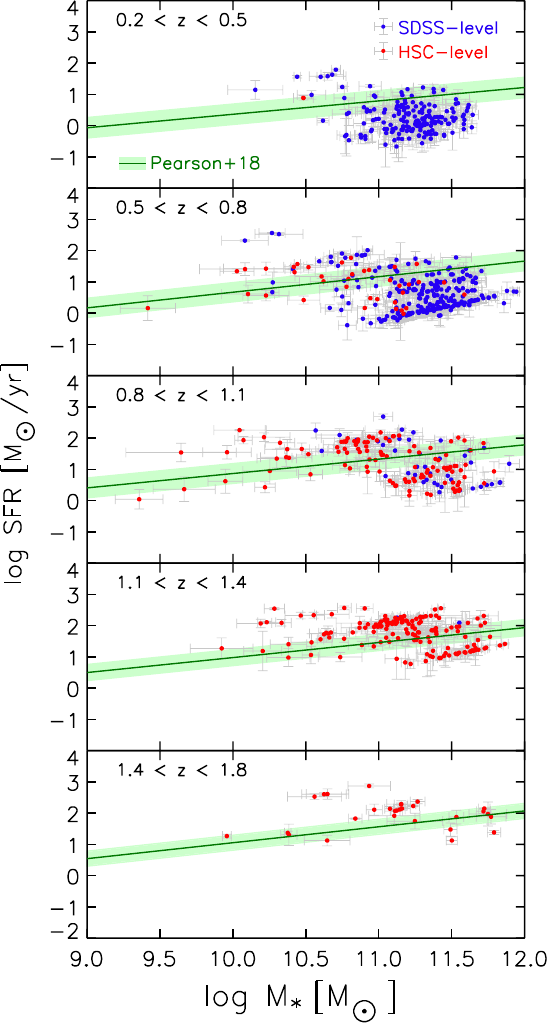}
\caption{Stellar mass and SFR for HSC--FIRST RGs as a function of redshift. Blue and red points are the SDSS- and HSC-level RGs, respectively. The green lines are the main sequences (MSs) of SF galaxies at each redshift range provided by \cite{Pearson}. The green shaded regions correspond to an intrinsic scatter of each green line.}
\label{M_SFR}
\end{figure}

Figure \ref{M_SFR} shows the stellar mass and SFR for HSC--FIRST RGs as a function of redshift.
The $M_{*}-$SFR relations of MS galaxies as a function of redshift are also plotted that are provided by \cite{Pearson}.
They measured stellar mass and SFR by using multi-wavelength data including UV to FIR.
They also employed {\tt CIGALE} to derive those quantities by assuming same SFH, SSP, and IMF as this work.
This is important to do a fair comparison because different assumptions of SFH, SSP, and IMF induces a systematic offset for stellar mass and (particularly) SFR \citep[e.g.,][]{Maraston_10}.

At low redshift (0.2 $< z <$ 0.8), the majority of the SDSS-level objects lie below the MSs indicating that they are passive galaxies, which is consistent with a classical view of RGs in the local Universe \citep{Best_12}
At intermediate redshift (0.8 $< z <$ 1.1) that is an overlapped redshift regime between SDSS- and HSC-level RGs, they are widely distributed on $M_{*}-$SFR plane; from passive, MS, to starburst galaxies.
We find that there is no clear difference between SDSS- and HSC- level RGs.
At high redshift (1.1 $< z <$ 1.7), the majority of the HSC-level RGs is located at MS although some HSC-level RGs lie above the MS of SF galaxies.
Eventually, we confirmed that our HSC--FIRST RG sample contains various populations, including classical passive RGs, and normal SF galaxies, and starburst galaxies.

\subsection{AGN luminosity and SFR relation as a function of redshift}
\label{D_LAGNSFR}

We investigate the relation between AGN and SF activity for HSC--FIRST RGs.
Many studies have demonstrated that AGN activity (e.g., AGN bolometric luminosity) correlates with SF activity (e.g., FIR luminosity) especially for luminous AGNs \citep[e.g.,][]{Netzer,Shao,Rosario,Stanley,Ueda}.
Although we already showed AGN luminosity and SFR for SDSS- and HSC-level RGs (see Figure \ref{phys}), we here investigate their relationship as a function of redshift in order to check there is a difference in SDSS- and HSC-level RGs.

\begin{figure}
\plotone{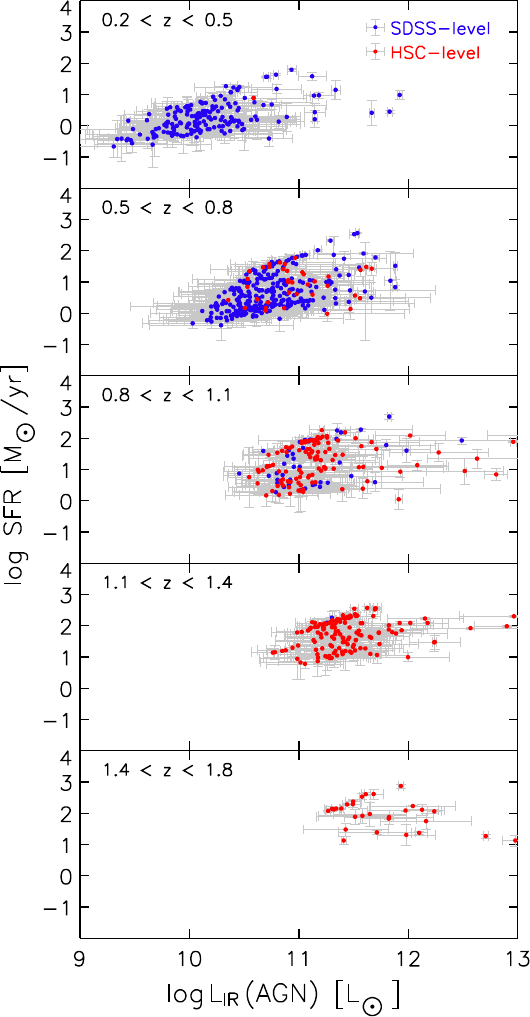}
\caption{The relationship between IR luminosity contributed from AGN and SFR for HSC--FIRST RGs as a function of redshift. Blue and red points are the SDSS- and HSC-level RGs, respectively.}
\label{LAGN_SFR}
\end{figure}

Figure \ref{LAGN_SFR} shows the relation between IR luminosity contributed from AGN, $L_{\rm IR}$ (AGN) and SFR as a function of redshift where $L_{\rm IR}$ (AGN) and SFR are derived in Section \ref{Sec_AGN} and \ref{S_SFR}, respectively.
We find that there is no clear difference in SDSS- and HSC-level RGs at a given redshift.

\subsection{Radio excess parameter}
\label{S_RE}

In section \ref{S_qir}, we found that $q_{\rm IR}$ of HSC--FIRST RGs is significantly lower than that of pure SF galaxies.
\cite{Del_Moro} defined ``radio-excess sources'' with $q_{\rm IR}$ $<$ 1.68 that corresponds to 3$\sigma$ deviation from the peak of the distribution for their sample.
According to their criterion, all of our HSC--FIRST RGs with TGSS data are radio-excess sources.
Even if we calculate $q_{\rm IR}$ for objects without TGSS data by adopting mean value of radio spectral index, about 98\% objects remain radio-excess sources.
Why are almost all HSC--FIRST RGs radio-excess sources?
We report this is due to our selection bias by comparing with much fainter RGs.

Here, we define ``radio-excess parameter'' that was introduced in \cite{Delvecchio};
\begin{equation}
q_{\rm excess} = \log \left(\frac{L_{\rm 1.4\,GHz}}{\rm SFR\,(IR)} \right),
\end{equation}
where $L_{\rm 1.4\, GHz}$ is what we obtained in Equation \ref{L14G}.
SFR (IR) is derived from IR luminosity contributed from SF in the same manner as \cite{Toba_17} \citep[see also][]{Kennicutt,Salim};
\begin{equation}
\label{IRSFR}
\log \, {\rm SFR}\, {\rm (IR)} =  \log \,L_{\rm IR}\, {\rm (SF)} - 9.966.
\end{equation}
\cite{Delvecchio} defined a threshold of radio excess sources as a function of redshift; if an object at a redshift $z$ has $q_{\rm excess} > 21.984 \times (1 + z)^{0.013}$, the object is classified as radio-excess source.
This definition is fairly consistent with that of \cite{Delvecchio}; radio-excess objects based on their selection satisfy the criterion of \cite{Delvecchio}, i.e., their $qi_{\rm IR}$ values are less than 1.68.

\begin{figure}
\plotone{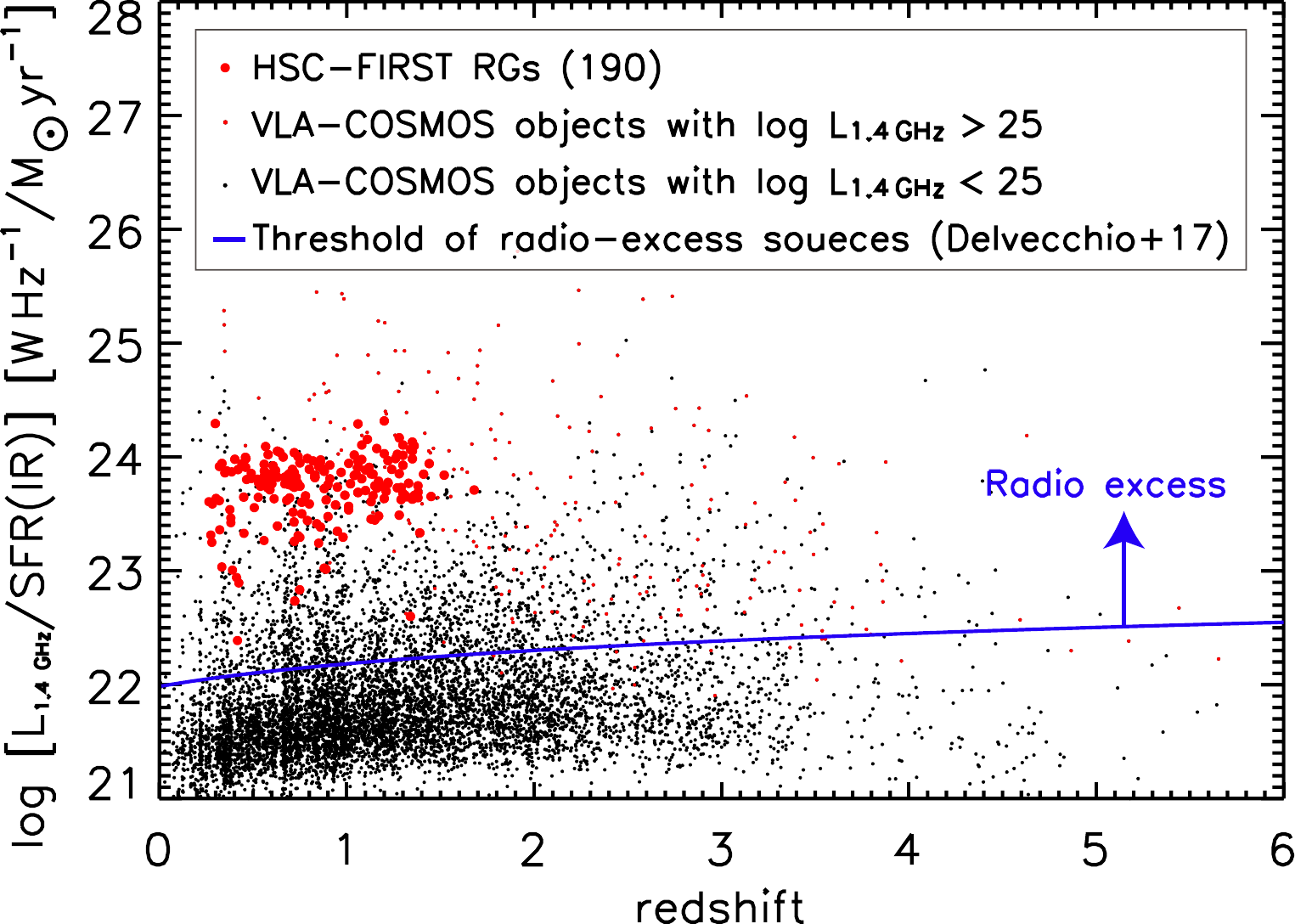}
\caption{$q_{\rm excess} = \log$ [$L_{\rm 1.4\,GHz}$/SFR\,(IR)] (radio excess parameter) as a function of redshift. Small dots are radio sources discovered by the VLA-COSMOS 3 GHz Large Project \citep{Smolcic_17a}. Large circles with red color are our HSC--FIRST RGs. Small red dots means VLA-COSMOS sources with $\log\, L_{\rm 1.4\,GHz} > 25$ W Hz$^{-1}$. Blue line is the threshold of radio excess sources that is formulated as $q_{\rm excess} = 21.984 \times (1 + z)^{0.013}$ \citep{Delvecchio}. Objects with $q_{\rm excess}$ greater than this threshold are classified as radio-excess sources.}
\label{excess}
\end{figure}

Figure \ref{excess} shows radio excess parameter as a function of redshift for HSC-FIRST objects with TGSS data.
Low luminosity radio sources found by VLA-COSMOS 3 GHz large project \citep{Smolcic_17a,Smolcic_17b} are also plotted.
We found that almost all RGs with $\log\,L_{\rm 1.4\,GHz} > 25.0$ W Hz$^{-1}$ are classified as radio-excess sources.
Since 156/190 ($\sim$82 \%) HSC--FIRST RG sample has $\log\,L_{\rm 1.4\,GHz} > 25.0$ W Hz$^{-1}$, we conclude that the fact that our sources are radio-excess objects may be due to the flux cut ($f_{\rm 1.4\,GHz} >$ 1.0 mJy) for HSC-FIRST RGs (see Section \ref{SSS}).
We also confirm that the origin of radio excess is due to AGNs that boost radio luminosity because their SFRs are basically normal SF galaxies (see Section \ref{S_SFR}).
Indeed, \cite{Del_Moro} reported that the fraction of radio-excess objects increases with X-ray luminosity.
They also found that roughly half of these radio-excess AGNs are not detected in the deep {\it Chandra} X-ray data.
Taking the fact that HSC-level objects have large $E(B-V)_*$ (see Section \ref{S_EBV}) into account, these results could indicate that particularly some HSC--level RGs harbor heavily obscured AGNs.

\subsection{Accretion rate}
\label{Edd}
We discuss the BH mass accretion rate of SDSS- and HSC-level objects.
RGs are classified into low-excitation RGs (LERGs) and high-excitation RGs (HERGs) based on their optical spectra \citep[e.g.,][]{Laing,Buttiglione}.
Many works studied on physical properties of LERGs and HERGs, and revealed that HERGs tend to have low stellar mass and high SFR  while LERGs tend to be reside in denser enlivenment \citep[e.g.,][]{Best_12,Janssen,Ching}.
In terms of {\it WISE} colors, LERGs are basically distributed at ellipticals/spirals/LIRGs while HERGs are basically distributed at Seyferts/starbursts/ULIRGs \citep{Gurkan_14,Yang,Mingo,Whittam}.
This result could indicate that the relation between LERGs and HERGs is likely to be similar as that of SDSS- and HSC-level objects \citep[e.g.,][]{Prescott}.
Because the observational characteristics of HERGs and LERGs are mainly driven by the accretion rate on to the SMBH \citep{Best_12}, it is expected that accretion rate of HSC-level objects would differ from SDSS-level objects.
 
First, we checked a difference of observational quantities; the ratio of rest-frame 22 and 3.4 $\micron$ in the same manner as \cite{Gurkan_14}.
Since the rest-frame 22 $\micron$ luminosity is a good tracer of AGN luminosity while rest-frame 3.4 $\micron$ luminosity roughly corresponds to stellar mass, their luminosity ratio is a proxy of the Eddington-scaled accretion rate.
Rest-frame 3.4 and 22 $\micron$ luminosities were derived from {\tt CIGALE} that conducted a convolution integral of best-fit SED with filter response functions of {\it WISE} W1 and W4 bands.

\begin{figure}
\plotone{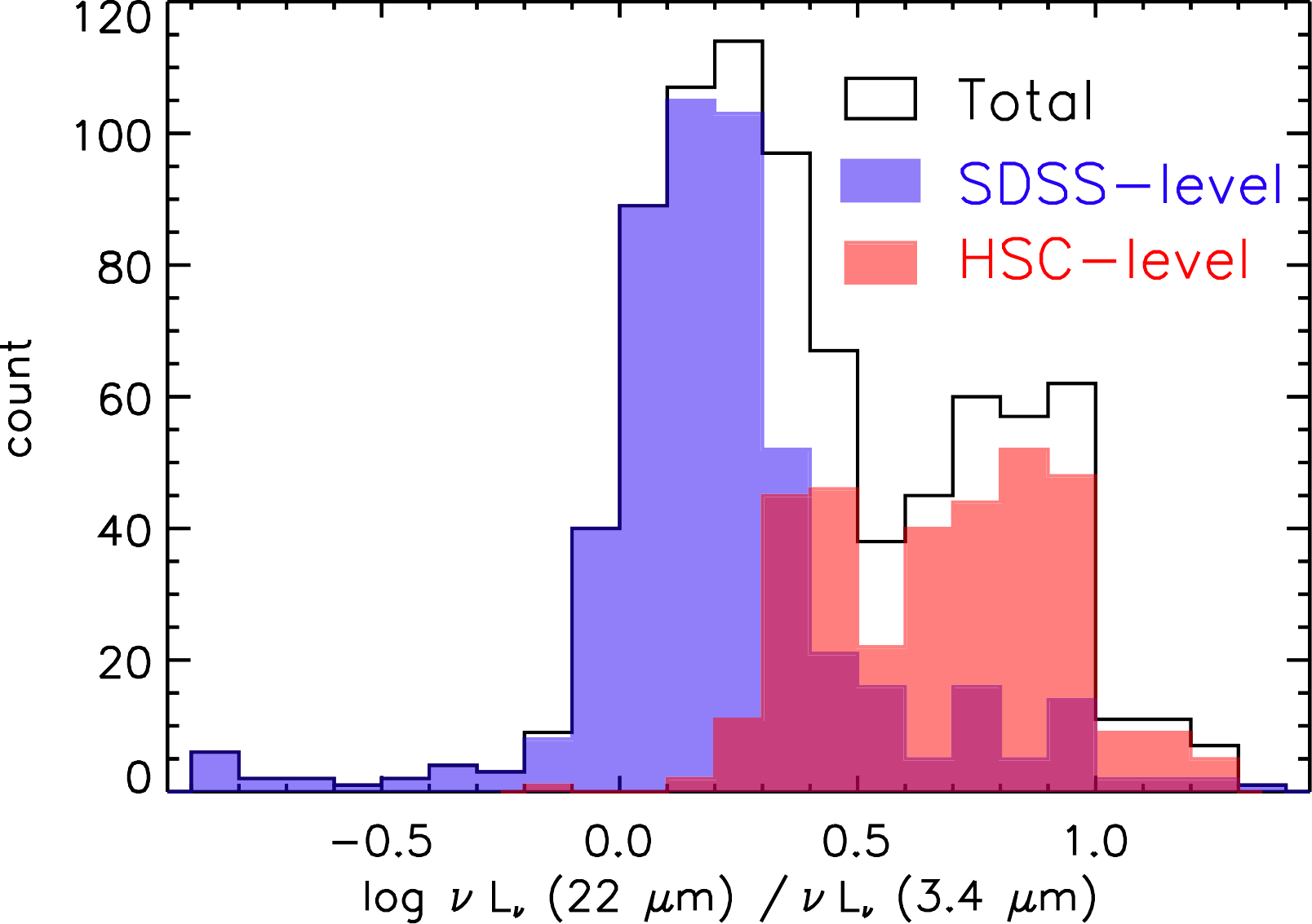}
\caption{Ratio of rest-frame 22 $\micron$ and 3.4 $\micron$ luminosities for SDSS-level (red), HSC-level (blue), and total (black) RGs.}
\label{L2234}
\end{figure}

Figure \ref{L2234} shows histogram of luminosity ratio of rest-frame 22 $\micron$ and 3.4 $\micron$ for HSC--FIRST RGs.
There is a clear difference between SDSS- and HSC-level objects; the luminosity ratio of HSC-level objects is systematically larger than that of SDSS-level objects.
This result suggests that HSC-level objects have a high Eddington-scaled accretion rate compared to SDSS-level objects.

Next, we performed a rough estimate of Eddington ratio ($\lambda_{\rm Edd}$) of our RG sample, in the same manner as \cite{Toba_17b} \citep[see also][]{Mingo,Whittam}.
The BH mass ($M_{\rm BH}$) was estimated from stellar mass by using an empirical relation with a scatter of 0.24 dex, reported in \cite{Reines};
\begin{equation}
\log \,(M_{\rm BH}/M_{\sun}) = 7.45 + 1.05 \times \log\,(M_*/10^{11}\,M_{\sun}),
\end{equation}
and we converted it to Eddington luminosity ($L_{\rm Edd}$).
The bolometric luminosity ($L_{\rm bol}$) is estimated by integrating the best-fit SED template of AGN component output by {\tt CIGALE} over wavelengths longward of Ly$\alpha$.

\begin{figure}
\plotone{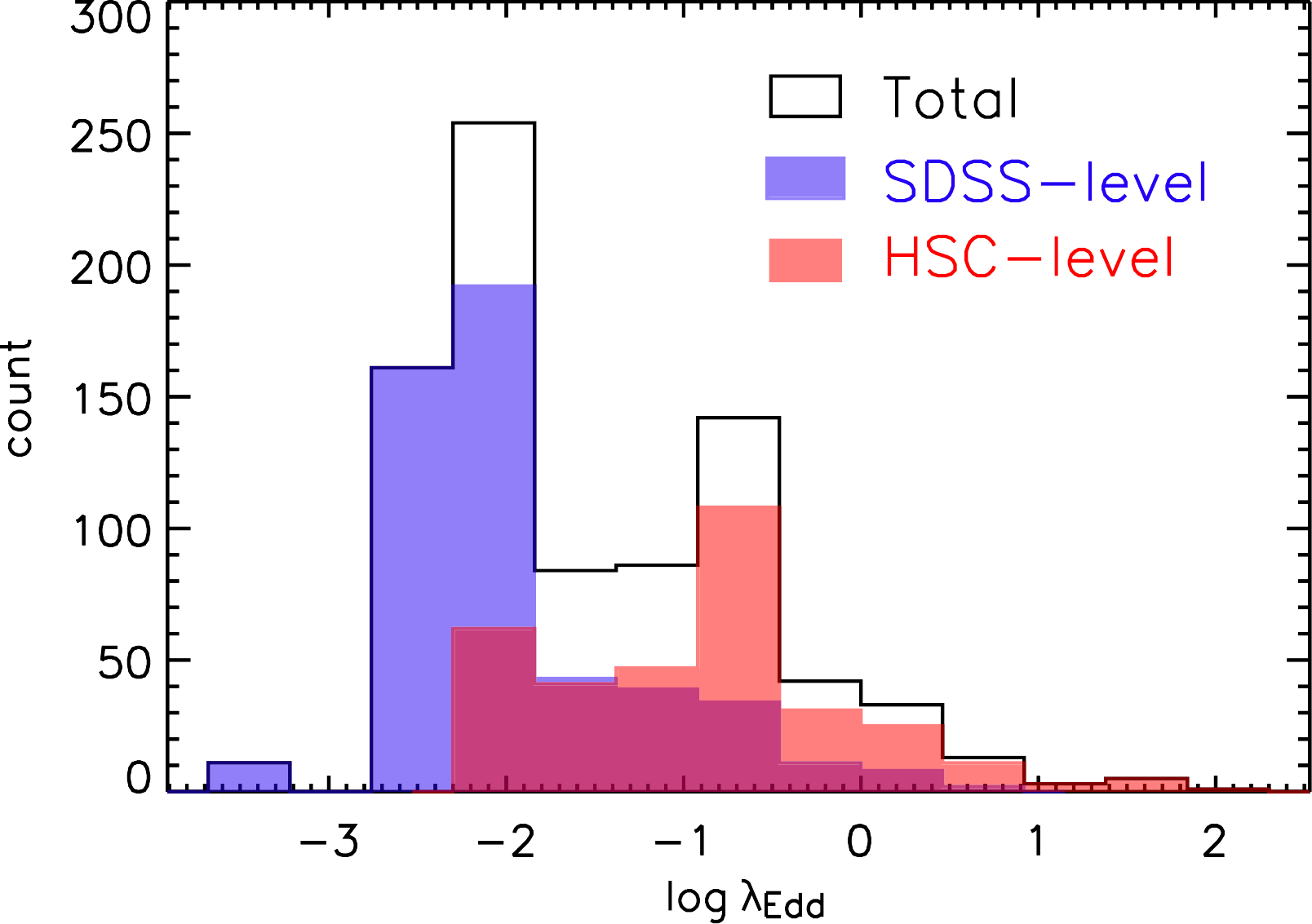}
\caption{Histogram of Eddington ratio for SDSS-level (red), HSC-level (blue), and total (black) RGs.}
\label{lEdd}
\end{figure}

Figure \ref{lEdd} shows histogram of $\lambda_{\rm Edd}$ (= $L_{\rm bol}/L_{\rm Edd})$ of HSC--FIRST RGs.
The HSC-level objects clearly have large Eddington ratio compared with SDSS-level objects.
The mean values of ($\lambda_{\rm Edd}$) of SDSS- and HSC-level objects are -1.95 and -0.94, respectively, indicating that HSC-level objects have actively growing SMBHs in their center.
We note that a fraction of HSC-level objects has $\lambda_{\rm Edd} > 1$.
Their Eddington ratios may be overestimated due to underestimate of their black hole mass.
We used the empirical relation of stellar mass and BH mass provided by \cite{Reines} that is optimized for broad line AGNs with $6 < \log \,(M_{\rm BH}/M_{\sun}) < 8$ at $z < 0.1$.
If we use an another empirical relation for elliptical galaxies provided by \cite{Reines}, the resultant BH masses are roughly one order of magnitude larger than those we reported above.
\cite{McLure} also reported that the BH to bulge mass ratio of radio-loud AGNs increases with incensing redshift; given a bulge mass of an object at $z > 1$, its BH mass is larger than that of local universe.
Since a large fraction of HSC-level objects is radio-loud AGNs at $z > 1$, their BH masses would be underestimated.
Nevertheless, the difference in Eddington ratio between SDSS- and HSC-level objects seems to be significant even if the BH mass of HSC-level objects would be underestimated by 0.5-1 dex.

\subsection{Duty cycle of the HSC--FIRST RGs}
\label{DC}
Finally, we briefly discuss the duty cycle of SDSS- and HSC- level RGs and their evolutionally link.
It should be noted that since our RG sample might be affected by systematic uncertainty and selection bias as discussed in Section \ref{SB} and \ref{PU}, the estimated duty cycle may also have a large uncertainty.
Nevertheless, it is worth discussing how the RG sample discovered by the HSC and FIRST can be interpreted in the framework of galaxy formation and evolution.
We selected 501 SDSS-level RGs at $ 0.1 < z < 1.2$ while 334 HSC-level RGs at $0.5 < z < 1.7$ in $\sim 94.7$ deg$^{2}$.
The corresponding co-moving volume density of SDSS- and HSC- level RGs is $8.9 \times 10^{-7}$ and $3.3 \times 10^{-7}$ cMpc$^{-3}$, respectively\footnote{cMpc is a co-moving distance in unit of Mpc.}.
On the other hand, the range of stellar mass derived by {\tt CIGALE} for SDSS-level RGs is $11.0 < \log \,(M_{*}/M_{\sun}) < 11.6$ while that for HSC-level RGs is $ 10.6 < \log \,(M_{*}/M_{\sun}) < 11.5$.
According to stellar mass function of massive galaxies provided by \cite{Kajisawa}\footnote{\cite{Kajisawa} assumes \cite{Salpeter} IMF. So, we re-calculated the volume density based on Chabrier IMF.}, the volume density of galaxies with same redshift and stellar mass range as SDSS- and HSC- level RGs is $3.1 \times 10^{-4}$ and $7.9 \times 10^{-4}$ cMpc$^{-3}$, respectively.
If we assume that massive galaxies with $\log \,(M_{*}/M_{\sun}) \sim 11.0$ have an experience of HSC--FIRST RG phase at least once during a redshift range in which they are observed (i.e., 6.95 and 4.82 Gyr for SDSS- and HSC-level RGs, respectively), the resultant duty cycle of SDSS- and HSC-level RGs is 0.003 ($\sim$ 19.6 Myr) and 0.0004 ($\sim$ 2.0 Myr), respectively.

Since the stellar mass of the vast majority of the optically faint RGs is indeed as massive as the bright RGs, there may be a possibility that they are evolutionally linked.
We may be witnessing short duty cycle phenomena, which may be able to quench SF activity at $z\sim$ 1.0 or keep quenching SF activity at $z\sim$ 0.5 and to activate AGNs in relatively massive galaxies.
On the other hand, the duty cycle of HSC-level RGs seems to be too short as a duration of radio jet activity in powerful RGs.
One possibility is that our assumption to derive the duty cycle (i.e., massive galaxies have an experience of HSC--FIRST RG phase at least once during their redshift range) is too strict.
If we assume that  massive galaxies would have an experience of HSC--FIRST RG phase at least once in the history of Universe, the duty cycle could be about 10 Myr.

\section{Summary}
\label{summary}

In this work, we investigated the physical properties of optically-faint RGs with $f_{\rm 1.4 GHz} >$ 1 mJy selected by HSC and FIRST, whose nature has been poorly understood so far.
We constructed a subsample of 1056 RGs with reliable redshift and multi-wavelength data from optical to radio, among a sample of 1943 RGs in $\sim$100 deg$^{2}$.
By conducting the SED fitting with {\tt CIGALE}, we obtained reliable physical quantities of 835 objects at $0 < z \leq 1.7$.
Thanks to the deep optical imaging with HSC, we are able to investigate physical quantities of luminous RGs even at $z > 0.5$ that cannot be probed by previous optical surveys.
We investigate the physical quantities as a function of redshift.
In addition, we discuss the physical difference between optically-bright, SDSS-level RGs with $i <$ 21.3 mag (mean $z$ = 0.57) and optically-faint, HSC-level RGs with $i \geq$ 21.3 mag (mean $z$ = 1.10).
We summarize resultant properties (mean value of each quantity for SDSS- and HSC-level RGs and total RG sample) in Table \ref{sum}.
The main results are as follows.
\begin{enumerate}
\item Color excess, $E(B-V)_{*}$, increases with increasing redshift, and thus $E(B-V)_{*}$ of HSC-level objects is larger than that of SDSS-level objects (Section \ref{S_EBV}).
\item Stellar mass is not significantly correlated with redshift. But the mean stellar mass of HSC-level objects is slightly smaller than that of SDSS-level objects.
On the other hand, SFR increases with increasing redshift, and thus SFR of HSC-level objects is larger than that of SDSS-level objects (Sections \ref{S_M} and \ref{S_SFR}).
\item Total IR luminosity and IR luminosity contributed from AGN increase with increasing redshift, and thus those luminosities of HSC-level objects are larger than those of SDSS-level objects. Most HSC-level objects are classified as ULIRGs with $\log \,(L_{\rm IR}/L_{\sun}) > 12.0$ (Sections \ref{Sec_AGN} and \ref{Sec_IR}).
\item Radio spectral index ($\alpha_{\rm radio}$) and luminosity ratio of IR and radio ($q_{\rm IR}$) do not significantly depend redshift. 
However, the mean $\alpha_{\rm radio}$ of HSC-level objects is slightly larger than that of SDSS-level objects while mean $q_{\rm IR}$ of HSC-level objects is smaller than that of SDSS-level objects (Sections \ref{Rindex} and \ref{S_qir}).
\item Eddington ratio ($\lambda_{\rm Edd}$) of HSC-level objects is larger than that of SDSS-level objects, suggesting that optically-faint HSC-level RGs discovered by HSC and FIRST could be dust-obscured AGNs with actively growing SMBHs (Section \ref{Edd}).
\end{enumerate}

\startlongtable
\begin{deluxetable}{lrrr}
\tablecaption{Summary of physical properties (the mean value of each quantity) of 835 HSC--FIRST RGs.\label{sum}}
\tablehead{
\colhead{Physical properties} & \colhead{SDSS-level} & \colhead{HSC-level} & \colhead{Total}
}
\startdata
$E(B-V)_{*}$							&	0.19 	&	0.45	&	0.30\\
$\log \,(M_{*}/M_{\sun})$				&	11.26	&	11.08	&	11.19\\
$\log$ SFR [$M_{*}$ yr$^{-1}$]			&	0.55	&	1.51	&	0.93\\
$\log \,[L_{\rm IR}$(AGN)/$L_{\sun}]$	&	10.56	&	11.32	&	10.87\\
$\log \,(L_{\rm IR}/L_{\sun})$			&	11.31	&	12.04	&	11.61\\
$\alpha_{\rm radio}$					&	0.72	&	0.74	&	0.73\\
$q_{\rm IR}$							&	0.37	&	0.31	&	0.34\\
$\log\, \lambda_{\rm Edd}$				&	-1.95	&	-0.94	&	-1.54\\
\enddata
\end{deluxetable}

Overall, our HSC-FIRST sample seems to have a variety of RGs including classical ones with massive host, low SFR and low Eddington ratio, and sort of new population with less massive host, high SFR and high Eddington ratio.
In particular, the later ones are optically-faint and high redshift RGs that cannot be discovered by the SDSS, whose properties differ from a classical view of RGs.
We conclude that the WERGS project with HSC and FIRST explores new population that would be missed by previous surveys.

\acknowledgments

We gratefully acknowledge the anonymous referee for a careful reading of the manuscript and very helpful comments.
We are deeply thankful to Prof. Denis Burgarella, Dr. M\'ed\'eric Boquien, and Dr. Laure Ciesla for helping us to understand {\tt CIGALE} code.

The Hyper Suprime-Cam (HSC) collaboration includes the astronomical communities of Japan and Taiwan,
and Princeton University. The HSC instrumentation and software were developed by the National Astronomical Observatory of Japan (NAOJ), the Kavli Institute for the Physics and Mathematics of the Universe (Kavli IPMU), the University of Tokyo, the High Energy Accelerator Research Organization (KEK), the Academia Sinica Institute for Astronomy and Astrophysics in Taiwan (ASIAA), and Princeton University. 
Funding was contributed by the FIRST program from Japanese Cabinet Office, the Ministry of Education, Culture, Sports, Science and Technology (MEXT), the Japan Society for the Promotion of Science (JSPS), Japan Science and Technology Agency (JST), the Toray Science Foundation, NAOJ, Kavli IPMU, KEK, ASIAA, and Princeton University. 

This paper makes use of software developed for the Large Synoptic Survey Telescope. We thank the LSST Project for making their code available as free software at \url{http://dm.lsstcorp.org}.

The Pan-STARRS1 Surveys (PS1) have been made possible through contributions of the Institute for Astronomy, the University of Hawaii, the Pan-STARRS Project Office, the Max-Planck Society and its participating institutes, the Max Planck Institute for Astronomy, Heidelberg and the Max Planck Institute for Extraterrestrial Physics, Garching, The Johns Hopkins University, Durham University, the University of Edinburgh, Queen's University Belfast, the Harvard-Smithsonian Center for Astrophysics, the Las Cumbres Observatory Global Telescope Network Incorporated, the National Central University of Taiwan, the Space Telescope Science Institute, the National Aeronautics and Space Administration under Grant No. NNX08AR22G issued through the Planetary Science Division of the NASA Science Mission Directorate, the National Science Foundation under Grant No. AST-1238877, the University of Maryland, and Eotvos Lorand University (ELTE).

Based on data collected at the Subaru Telescope and retrieved from the HSC data archive system, which is operated by Subaru Telescope and Astronomy Data Center at National Astronomical Observatory of Japan.

The National Radio Astronomy Observatory is a facility of the National Science Foundation operated under cooperative agreement by Associated Universities, Inc.

Based on data products from observations made with ESO Telescopes at the La Silla Paranal Observatory under programme IDs 177.A-3016, 177.A-3017 and 177.A-3018, and on data products produced by Target/OmegaCEN, INAF-OACN, INAF-OAPD and the KiDS production team, on behalf of the KiDS consortium. OmegaCEN and the KiDS production team acknowledge support by NOVA and NWO-M grants. Members of INAF-OAPD and INAF-OACN also acknowledge the support from the Department of Physics \& Astronomy of the University of Padova, and of the Department of Physics of Univ. Federico II (Naples).

This publication has made use of data from the VIKING survey from VISTA at the ESO Paranal Observatory, programme ID 179.A-2004. Data processing has been contributed by the VISTA Data Flow System at CASU, Cambridge and WFAU, Edinburgh.

The Herschel-ATLAS is a project with Herschel, which is an ESA space observatory with science instruments provided by European-led Principal Investigator consortia and with important participation from NASA. The H-ATLAS website is \url{http://www.h-atlas.org/}.

We thank the staff of the GMRT that made these observations possible. GMRT is run by the National Centre for Radio Astrophysics of the Tata Institute of Fundamental Research.

This research has made use of the NASA/ IPAC Infrared Science Archive, which is operated by the Jet Propulsion Laboratory, California Institute of Technology, under contract with the National Aeronautics and Space Administration.

Funding for SDSS-III has been provided by the Alfred P. Sloan Foundation, the Participating Institutions, the National Science Foundation, and the U.S. Department of Energy Office of Science. The SDSS-III web site is http://www.sdss3.org/.
SDSS-III is managed by the Astrophysical Research Consortium for the Participating Institutions of the SDSS-III Collaboration including the University of Arizona, the Brazilian Participation Group, Brookhaven National Laboratory, Carnegie Mellon University, University of Florida, the French Participation Group, the German Participation Group, Harvard University, the Instituto de Astrofisica de Canarias, the Michigan State/Notre Dame/JINA Participation Group, Johns Hopkins University, Lawrence Berkeley National Laboratory, Max Planck Institute for Astrophysics, Max Planck Institute for Extraterrestrial Physics, New Mexico State University, New York University, Ohio State University, Pennsylvania State University, University of Portsmouth, Princeton University, the Spanish Participation Group, University of Tokyo, University of Utah, Vanderbilt University, University of Virginia, University of Washington, and Yale University.

GAMA is a joint European-Australasian project based around a spectroscopic campaign using the Anglo-Australian Telescope. The GAMA input catalogue is based on data taken from the Sloan Digital Sky Survey and the UKIRT Infrared Deep Sky Survey. Complementary imaging of the GAMA regions is being obtained by a number of independent survey programmes including GALEX MIS, VST KiDS, VISTA VIKING, WISE, Herschel-ATLAS, GMRT and ASKAP providing UV to radio coverage. GAMA is funded by the STFC (UK), the ARC (Australia), the AAO, and the participating institutions. The GAMA website is http://www.gama-survey.org/ .

Y.Toba and W.H.Wang acknowledge the support from the Ministry of Science and Technology of Taiwan (MOST 105-2112-M-001-029-MY3). 
K.Ichikawa is supported by Program for Establishing a Consortium for the Development of Human Resources in Science and Technology, Japan Science and Technology Agency (JST).
This work is supported by JSPS KAKENHI Grant numbers 18J01050 and 19K14759 (Y.Toba), 16H01101, 16H03958, and 17H01114 (T.Nagao), and 18K13584 (K.Ichikawa), and 17K05384 (Y.Ueda).

Numerical computations/simulations were carried out (in part) using the SuMIRe cluster operated by the Extragalactic OIR group at ASIAA.

\vspace{5mm}

\facilities{Subaru (HSC), VST, ESO:VISTA, WISE, Herschel, VLA, GMRT, IRSA}

\software{IDL, IDL Astronomy User's Library \citep{Landsman}, TOPCAT \citep{Taylor}, {\tt CIGALE} \citep{Boquien}}

\appendix
\section{HSC-FIRST radio galaxy catalog}
\label{app1}

We provide HSC--FIRST RG catalog that includes 1056 RGs used for the SED fitting with {\tt CIGALE}.
The catalog description is summarized in Table \ref{catalog}.

\startlongtable
\begin{deluxetable}{lccl}
\tablecaption{Format and column descriptions of HSC--FIRST RG catalog.\label{catalog}}
\tablehead{
\colhead{Column name} & \colhead{Format} & \colhead{Unit} & \colhead{Description}
}
\startdata
ID	&	LONG	&	&	unique id \\
Name & STRING 	& 	& 	object name in the FIRST catalog \\
R.A.	&	DOUBLE	&	degree	& Right Assignation (J2000.0) from HSC S16a wide catalog\\
Decl.	&	DOUBLE	&	degree	& Declination (J2000.0) from HSC S16a wide catalog\\
Redshift	&	DOUBLE & 	& Redshift \\
Ref\_redshift & STRING & & Reference of redshift (mizuki/SDSS-DR12/GAMA-DR2/WIGGLEZ-DR2) \\
$u$mag			&	DOUBLE & AB mag. & $u$-band magnitude from KiDS DR3\\
$u$mag\_err		&	DOUBLE & AB mag. & $u$-band magnitude error from KiDS DR3\\
$g$mag			&	DOUBLE & AB mag. & $g$-band magnitude from HSC S16a wide catalog\\
$g$mag\_err		&	DOUBLE & AB mag. & $g$-band magnitude error from HSC S16a wide catalog\\
$r$mag			&	DOUBLE & AB mag. & $r$-band magnitude from HSC S16a wide catalog\\
$r$mag\_err		&	DOUBLE & AB mag. & $r$-band magnitude error from HSC S16a wide catalog\\
$i$mag			&	DOUBLE & AB mag. & $i$-band magnitude from HSC S16a wide catalog\\
$i$mag\_err		&	DOUBLE & AB mag. & $i$-band magnitude error from HSC S16a wide catalog\\
$z$mag			&	DOUBLE & AB mag. & $z$-band magnitude from HSC S16a wide catalog\\
$z$mag\_err		&	DOUBLE & AB mag. & $z$-band magnitude error from HSC S16a wide catalog\\
$y$mag			&	DOUBLE & AB mag. & $y$-band magnitude from HSC S16a wide catalog\\
$y$mag\_err		&	DOUBLE & AB mag. & $y$-band magnitude error from HSC S16a wide catalog\\
$j$mag			&	DOUBLE & AB mag. & $j$-band magnitude from VIKING DR3\\
$j$mag\_err		&	DOUBLE & AB mag. & $j$-band magnitude error from VIKING DR3\\
$h$mag			&	DOUBLE & AB mag. & $h$-band magnitude from VIKING DR3\\
$h$mag\_err		&	DOUBLE & AB mag. & $h$-band magnitude error from VIKING DR3\\
$k$smag			&	DOUBLE & AB mag. & $k$s-band magnitude from VIKING DR3\\
$k$smag\_err	&	DOUBLE & AB mag. & $k$s-band magnitude error from VIKING DR3\\
w1mag			&	DOUBLE & Vega mag & 3.4 $\micron$ magnitude from ALLWISE\\
w1mag\_err		&	DOUBLE & Vega mag & 3.4 $\micron$ magnitude error from ALLWISE\\
w2mag			&	DOUBLE & Vega mag & 4.6 $\micron$ magnitude from ALLWISE\\
w2mag\_err		&	DOUBLE & Vega mag & 4.6 $\micron$ magnitude error from ALLWISE\\
w3mag			&	DOUBLE & Vega mag & 12 $\micron$ magnitude from ALLWISE\\
w3mag\_err		&	DOUBLE & Vega mag & 12 $\micron$ magnitude error from ALLWISE\\
w4mag			&	DOUBLE & Vega mag & 22 $\micron$ magnitude from ALLWISE\\
w4mag\_err		&	DOUBLE & Vega mag & 22 $\micron$ magnitude error from ALLWISE\\
A\_u			&	DOUBLE &	mag		& Galactic extinction correction for $u$-band\\
A\_g			&	DOUBLE &	mag		& Galactic extinction correction for $g$-band\\
A\_r			&	DOUBLE &	mag		& Galactic extinction correction for $r$-band\\
A\_i			&	DOUBLE &	mag		& Galactic extinction correction for $i$-band\\
A\_z			&	DOUBLE &	mag		& Galactic extinction correction for $z$-band\\
A\_y			&	DOUBLE &	mag		& Galactic extinction correction for $y$-band\\
A\_j			&	DOUBLE &	mag		& Galactic extinction correction for $j$-band\\
A\_h			&	DOUBLE &	mag		& Galactic extinction correction for $h$-band\\
A\_ks			&	DOUBLE &	mag		& Galactic extinction correction for $k$s-band\\
Flux\_34		&	DOUBLE & mJy	& Flux density at 3.4 $\micron$ \\	
Flux\_34\_err	&	DOUBLE & mJy	& Uncertainty of flux density at 3.4 $\micron$ \\
Flux\_46		&	DOUBLE & mJy	& Flux density at 4.6 $\micron$ \\	
Flux\_46\_err	&	DOUBLE & mJy	& Uncertainty of flux density at 4.6 $\micron$ \\
Flux\_12		&	DOUBLE & mJy	& Flux density at 12 $\micron$ \\	
Flux\_12\_err	&	DOUBLE & mJy	& Uncertainty of flux density at 12 $\micron$ \\
Flux\_22		&	DOUBLE & mJy	& Flux density at 22 $\micron$ flux density \\	
Flux\_22\_err	&	DOUBLE & mJy	& Uncertainty of flux density at 22 $\micron$ \\
Flux\_100		&	DOUBLE & mJy	& Flux density at 100 $\micron$ from H-ATLAS DR1\\
Flux\_100\_err	&	DOUBLE & mJy	& Uncertainty of flux density at 100 $\micron$ from H-ATLAS DR1\\
Flux\_160		&	DOUBLE & mJy	& Flux density at 160 $\micron$ from H-ATLAS DR1\\
Flux\_160\_err	&	DOUBLE & mJy	& Uncertainty of flux density at 160 $\micron$ from H-ATLAS DR1\\
Flux\_250		&	DOUBLE & mJy	& Flux density at 250 $\micron$ from H-ATLAS DR1\\
Flux\_250\_err	&	DOUBLE & mJy	& Uncertainty of flux density at 250 $\micron$ from H-ATLAS DR1\\
Flux\_350		&	DOUBLE & mJy	& Flux density at 350 $\micron$ from H-ATLAS DR1\\
Flux\_350\_err	&	DOUBLE & mJy	& Uncertainty of flux density at 350 $\micron$ from H-ATLAS DR1\\
Flux\_500		&	DOUBLE & mJy	& Flux density at 500 $\micron$ from H-ATLAS DR1\\
Flux\_500\_err	&	DOUBLE & mJy	& Uncertainty of flux density at 500 $\micron$ from H-ATLAS DR1\\
Flux\_14G		&	DOUBLE & mJy	& Flux density at 1.4 GHz from FIRST\\
Flux\_14G\_err	&	DOUBLE & mJy	& Uncertainty pf flux density at 1.4 GHz from FIRST\\
Flux\_150M		&	DOUBLE & mJy	& Flux density at 150 MHz from TGSS ADR1\\
Flux\_150M\_err	&	DOUBLE & mJy	& Uncertainty pf flux density at 150 MHz from TGSS ADR1\\
Flag\_u			&	INT	&& Flag for $u$-band data (0: CIGALE. 1: CIGALE with 3$\sigma$ upper limit, 2: non CIGALE\\
Flag\_g			&	INT	&& Flag for $g$-band data (0: CIGALE. 1: CIGALE with 3$\sigma$ upper limit, 2: non CIGALE\\
Flag\_r			&	INT	&& Flag for $r$-band data (0: CIGALE. 1: CIGALE with 3$\sigma$ upper limit, 2: non CIGALE\\
Flag\_i			&	INT	&& Flag for $i$-band data (0: CIGALE. 1: CIGALE with 3$\sigma$ upper limit, 2: non CIGALE\\
Flag\_z			&	INT	&& Flag for $z$-band data (0: CIGALE. 1: CIGALE with 3$\sigma$ upper limit, 2: non CIGALE\\
Flag\_y			&	INT	&& Flag for $y$-band data (0: CIGALE. 1: CIGALE with 3$\sigma$ upper limit, 2: non CIGALE\\
Flag\_j			&	INT	&& Flag for $j$-band data (0: CIGALE. 1: CIGALE with 3$\sigma$ upper limit, 2: non CIGALE\\
Flag\_h			&	INT	&& Flag for $h$-band data (0: CIGALE. 1: CIGALE with 3$\sigma$ upper limit, 2: non CIGALE\\
Flag\_ks		&	INT	&& Flag for $k$s-band data (0: CIGALE. 1: CIGALE with 3$\sigma$ upper limit, 2: non CIGALE\\
Flag\_34	&	INT	&& Flag for 3.4 $\micron$ data (0: CIGALE. 1: CIGALE with 3$\sigma$ upper limit, 2: non CIGALE\\
Flag\_46	&	INT	&& Flag for 4.6 $\micron$ data (0: CIGALE. 1: CIGALE with 3$\sigma$ upper limit, 2: non CIGALE\\
Flag\_12	&	INT	&& Flag for 12 $\micron$ data (0: CIGALE. 1: CIGALE with 3$\sigma$ upper limit, 2: non CIGALE\\
Flag\_22	&	INT	&& Flag for 22 $\micron$ data (0: CIGALE. 1: CIGALE with 3$\sigma$ upper limit, 2: non CIGALE\\
Flag\_100	&	INT	&& Flag for 100 $\micron$ data (0: CIGALE. 1: CIGALE with 3$\sigma$ upper limit, 2: non CIGALE\\
Flag\_160	&	INT	&& Flag for 160 $\micron$ data (0: CIGALE. 1: CIGALE with 3$\sigma$ upper limit, 2: non CIGALE\\
Flag\_250	&	INT	&& Flag for 250 $\micron$ data (0: CIGALE. 1: CIGALE with 3$\sigma$ upper limit, 2: non CIGALE\\
Flag\_350	&	INT	&& Flag for 350 $\micron$ data (0: CIGALE. 1: CIGALE with 3$\sigma$ upper limit, 2: non CIGALE\\
Flag\_500	&	INT	&& Flag for 500 $\micron$ data (0: CIGALE. 1: CIGALE with 3$\sigma$ upper limit, 2: non CIGALE\\
Flag\_14G	&	INT	&& Flag for 1.4 GHz data (0: CIGALE. 1: CIGALE with 3$\sigma$ upper limit, 2: non CIGALE\\
Flag\_150M	&	INT	&& Flag for 150 MHz data (0: CIGALE. 1: CIGALE with 3$\sigma$ upper limit, 2: non CIGALE\\
log\_L14G\tablenotemark{a}		&	DOUBLE & W Hz$^{-1}$ & Rest-frame luminosity density at 1.4 GHz\\
log\_L14G\_err	&	DOUBLE & W Hz$^{-1}$ & Uncertainty of rest-frame luminosity density at 1.4 GHz\\
E\_BV		&	DOUBLE	&	&	Color excess ($E(B-V)$) derived from {\tt CIGALE} \\
E\_BV\_err		&	DOUBLE	&	&	Uncertainty of color excess ($E(B-V)$) derived from {\tt CIGALE} \\
log\_M		& DOUBLE & $M_{\sun}$ & Stellar mass derived from {\tt CIGALE} \\
log\_M\_err 	& DOUBLE & $M_{\sun}$ & Uncertainty of stellar mass derived from {\tt CIGALE} \\
log\_SFR		& DOUBLE & $M_{\sun}$ yr$^{-1}$ & SFR derived from {\tt CIGALE} \\
log\_SFR\_err 	& DOUBLE & $M_{\sun}$ yr$^{-1}$ & Uncertainty of SFR derived from {\tt CIGALE} \\
log\_SFR\_IR		& DOUBLE & $M_{\sun}$ yr$^{-1}$ & SFR derived from Equation \ref{IRSFR} \\
log\_LIR			& DOUBLE & $L_{\sun}$ & IR luminosity derived from {\tt CIGALE} \\
log\_LIR\_err	& DOUBLE & $L_{\sun}$ & Uncertainty of IR luminosity derived from {\tt CIGALE} \\
log\_LIR\_AGN		& DOUBLE & $L_{\sun}$ & IR luminosity contributed from AGN derived from {\tt CIGALE} \\
log\_LIR\_AGN\_err	& DOUBLE & $L_{\sun}$ & Uncertainty of IR luminosity contributed from AGN derived from {\tt CIGALE} \\
alpha\_radio		& DOUBLE & & Radio spectral index ($\alpha_{\rm radio}$) derived from Equation \ref{index}\\
alpha\_radio\_err	& DOUBLE & & Uncertainty of radio spectral index ($\alpha_{\rm radio}$) \\
qir					& DOUBLE & & $q_{\rm IR}$ derived from Equation \ref{qir_eq}\\
qir\_err			& DOUBLE & & Uncertainty of $q_{\rm IR}$\\
DOF		&	INT		&	& Degree of freedom for the SED fitting\\
rechi2	&	DOUBLE	&	& Reduced $\chi^{2}$ derived from {\tt CIGALE}\\
log\_MBH		&	DOUBLE	&	$M_{\sun}$	&	Black hole mass derived from stellar mass\\
log\_Lbol	&	DOUBLE	&	$L_{\sun}$	&	Bolometric luminosity derived from the best-fit SED \\
log\_ledd & DOUBLE & & Eddington radio ($\lambda_{\rm Edd}$) \\
\enddata
\tablenotetext{a}{If an object has $\alpha_{\rm radio}$, $\log \, L_{\rm 1.4\, GHz}$ is derived from Equation \ref{L14G}. Otherwise, we assume $\alpha_{\rm radio}$ = 0.7 for a calculation (see Section \ref{SzLM}).}
\tablecomments{This table is available in its entirety in a machine-readable form in the online journal.}
\end{deluxetable}

\section{Best-fit SED for each radio galaxy}
\label{app2}
The best-fit SED derived by {\tt CIGALE} is available in Table \ref{template}.
We strongly encourage to use a template of objects with reduced $\chi^{2} <$ 5.0 for science.
In addition, if you use radio part of the best-fit SED, we recommend to employ the template only for objects with TGSS detections.

\startlongtable
\begin{deluxetable*}{lccl}
\tablecaption{Best-fit SED template of each HSC--FIRST RG.\label{template}}
\tablehead{
\colhead{Column name} & \colhead{Format} & \colhead{Unit} & \colhead{Description}
}
\startdata
ID			&	LONG	&			& unique id \\
Wavelength 	& 	DOUBLE 	& $\micron$ & wavelength \\
FNU			&	DOUBLE	& mJy		& flux density at each wavelength\\
LNU			&	DOUBLE	& W			& luminosity density at each wavelength\\
\enddata
\tablecomments{This table is available in its entirety in a machine-readable form in the online journal.}
\end{deluxetable*}


\end{document}